\documentclass[]{jfm}

\usepackage{graphicx}
\usepackage{amsmath,amssymb}
\usepackage{newtxtext}
\usepackage{newtxmath}
\usepackage{natbib}
\usepackage[colorlinks, citecolor=blue]{hyperref}

\newcommand{\RomanNumeralCaps}[1]
\linenumbers

\usepackage{mathtools}
\usepackage[dvipsnames]{xcolor}

\definecolor{colorA}{rgb}{0.12156862745098039,0.4666666666666667,0.7058823529411765}
\definecolor{colorB}{rgb}{1.0,0.4980392156862745,0.054901960784313725}
\definecolor{colorC}{rgb}{0.17254901960784313,0.6274509803921569,0.17254901960784313}
\definecolor{colorD}{rgb}{0.8392156862745098,0.15294117647058825,0.1568627450980392}
\definecolor{colorE}{rgb}{0.5803921568627451,0.403921568627451,0.7411764705882353}
\definecolor{colorF}{rgb}{0.5490196078431373,0.33725490196078434,0.29411764705882354}
\definecolor{colorG}{rgb}{0.5019607843137255,0.5019607843137255,0.5019607843137255}
\definecolor{colorH}{rgb}{0.0,0.0,0.0}

\usepackage{tikz}
\usepackage{pgfplots}

\usepackage[normalem]{ulem}
\usepackage{color}

\usepackage[textwidth=\dimexpr\textwidth-2cm\relax]{todonotes}
\makeatletter
\@mparswitchfalse%
\makeatother
\normalmarginpar 

\newcommand{\oo}{\color{black} \normalfont}
\newcommand{\bb}{\color{black} \normalfont}

\newcommand{\vs}{\color{black} \normalfont}

\newcommand{\hlrev}[1]{{\color{black}#1}} 

\title{Bursting bubble in an elasto-viscoplastic medium}

\author{
Arivazhagan~G.~Balasubramanian\aff{1,2}\corresp{\email{argb@mech.kth.se}},
Vatsal Sanjay\aff{3},
Maziyar Jalaal\aff{4}, 
Ricardo~Vinuesa\aff{1,2}
\and Outi Tammisola \aff{1,2}\corresp{\email{outi@mech.kth.se}}
}

\affiliation{
\aff{1}FLOW, Engineering Mechanics, KTH Royal Institute of Technology, SE-100 44 Stockholm, Sweden
\aff{2}Swedish e-Science Research Centre (SeRC), Stockholm, Sweden
\aff{3} Physics of Fluids Group, Max Planck Center for Complex Fluid Dynamics, and J.M. Burgers Center for Fluid Dynamics, University of Twente, P.O. Box 217, 7500AE Enschede,  The Netherlands
\aff{4} Van der Waals–Zeeman Institute, Institute of Physics, University of Amsterdam, 1098XH Amsterdam, The Netherlands
}

\begin{document}

\maketitle

\begin{abstract}


A gas bubble sitting at a liquid-gas interface can burst following the rupture of the thin liquid film separating it from the ambient, owing to the large surface energy of the resultant cavity. This bursting bubble forms capillary waves, a Worthington jet, and subsequent droplets for a Newtonian liquid medium.
However, rheological properties of the liquid medium like elasto-viscoplasticity can greatly affect these dynamics. Using direct numerical simulations, this study exemplifies how the complex interplay between elasticity (in terms of elastic stress relaxation) and yield stress influences the transient interfacial phenomena of bursting bubbles. We investigate how bursting dynamics depends on capillary, elastic, and yield stresses by exploring the parameter space of the Deborah number $\text{\De}$ (dimensionless relaxation time of elastic stresses) and the plastocapillary number $\mathcal{J}$ (dimensionless yield-stress of the medium), delineating four distinct characteristic behaviours. Overall, we observe a non-monotonic effect of elastic stress relaxation on the jet development while plasticity of the elasto-viscoplastic medium is shown to affect primarily the jet evolution only at faster relaxation times~(low $\text{\De}$). The role of elastic stresses on jet development is elucidated with the support of energy budgets identifying different modes of energy transfer within the elasto-viscoplastic medium.
The effects of elasticity on the initial progression of capillary waves and droplet formation are also studied. In passing, we study the effects of solvent-polymer-viscosity ratio on bursting dynamics and show that polymer viscosity can increase the jet thickness apart from reducing the maximum height of the jet.

\end{abstract}

\section{Introduction}
Bursting bubbles at a liquid-gas interface is a fundamental hydrodynamic process that has piqued the interest in various fields across multiple scales ranging from food processing industry \citep{woodcock1953, macintyre1972} to oceanic wave breaking \citep{veron2015, blanco2020, deike2022}. 
A typical daily realisation of bubble bursting occurs in a glass of champagne or other sparkling wine and is often credited for enhancing the mouthfeel of the taster \citep{ghabache2014, ghabache2016evap, seon2017}.  
Bubble bursting also plays a wider role in liquid fragmentation~\citep{villermaux2020fragmentation} and serves as a significant mechanism in facilitating the mass transfer of substances across the liquid-gas interface including
transporting pathogens from contaminated water \citep{blanchard1970, poulain2018, bourouiba2021, ji2022}.
The interactions of such bubbles with complex rheological fluids abound in nature. For example, the elastic and plastic fluid properties govern and influence pathogen transmission, and the pathogens might even adapt to or manipulate the chemical properties of the carrier fluids to benefit their own transmission \citep{bourouiba2021fluid}. Additionally, the presence of contaminants, surfactants, and oils in the marine boundary layer alters the bursting phenomenon and thereby affects the production of fine marine spray~\citep{ji2021, pierre2022influence, neel2022velocity, ji2023}. The rheological response of food products \citep{ahmed2016advances, mathijssen2023culinary} exemplifies yet another instance of the importance of understanding the mechanisms of bubble bursting in rheologically complex fluids. Such an understanding will also improve our knowledge of other natural phenomena, such as volcanic eruptions and underwater gas seep \citep{gonnermann2007fluid}.

\begin{figure}
	\centering
	\resizebox*{0.8\textwidth}{!}{\includegraphics{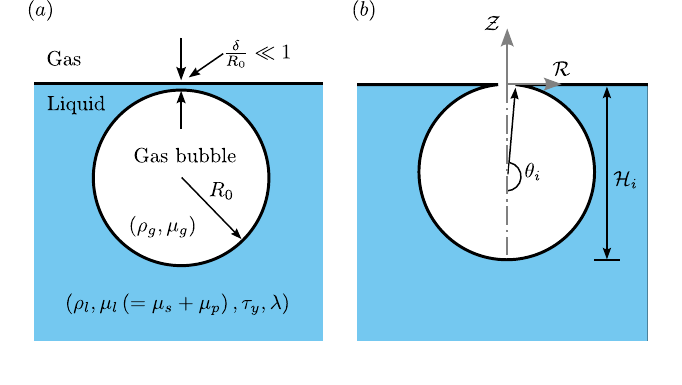}}
	\caption[]{Initial condition for bubble bursting: $(a)$ Gas bubble approaches the free interface forming a film of thickness $\delta$, $(b)$ The thin film is removed, resulting in a bubble cavity which is considered as the initial condition in our simulations.
	} 
	\label{fig_initial_condition}
\end{figure}

Unlike the rheologically simpler Newtonian fluids, elastic and plastic properties of the complex fluid govern the bubble bursting in addition to other factors such as buoyancy, surface tension, and viscosity. As an air bubble rises within the surrounding medium due to difference in density and approaches the liquid-air interface~(figure~\ref{fig_initial_condition}$a$), the thin liquid film gradually drains~\citep{allan1961} and ruptures subsequently, resulting in the formation of an open cavity~(as illustrated in figure~\ref{fig_initial_condition}$b$)~\citep{mason1954,zhang2013}. The open cavity is unstable due to large surface energy. It thereby collapses, leading to a sequence of dynamic events, including the propagation of capillary waves, which can potentially result in a Worthington jet~\citep{gekle2010,gordillo2019}. The Worthington jet may destabilise due to the Rayleigh--Plateau instability, leading to the formation of small droplets~\citep{gordillo2010,ghabache2014}.

Early studies bubble bursting began with a combination of experimental investigations and theoretical analyses, laying the foundation for identifying the underlying physics of the bursting mechanism in Newtonian fluids. With the progress in direct numerical simulation of multiphase flows~\citep{popinet2003,popinet2009},~\citet{deike2018} provided quantitative cross-validation of numerical and experimental studies. Further studies through a combination of experimental, numerical, and theoretical predictions~\citep{duchemin2002,walls2015,berny2020} revealed that the formation of droplets from the jet in the bubble bursting process is primarily determined by the viscous and gravitational effects.

On the other hand, the behaviour of bubble bursting in a different rheological medium has received less attention. Very recently, researchers have focused on studying the behaviour of such bubbles in non-Newtonian fluids~\citep{vatsal2021,ji2023,rodriguez2023,dixit2024}, as they exhibit unique flow characteristics that can significantly affect the bursting dynamics. Notably, \cite{rodriguez2023} explored the phenomenon of bubble-bursting in the presence of polymeric molecules, accounting for elastic effects induced by polymers. Their study revealed that the droplet emission is hindered due to extensional thickening in the jet. Another class of non-Newtonian fluids, called yield-stress fluids, exhibits a combination of solid and fluid behaviours. A thorough description and review of yield-stress fluids can be found in \textit{e.g.}~\cite{balmforth2014}. Within the category of yield-stress fluids, a specific type known as viscoplastic fluid acts as a rigid solid below the yield stress and flows like a viscous fluid when the shear stress exceeds the material's yield stress. The capillary flows of viscoplastic fluids have been studied for various droplets~\cite{jalaal2021spreading,van2023viscoplastic} and bubble problems~\cite{jalaal2016long,shemilt2023surfactant,shemilt2022surface,pourzahedi2022flow}. \cite{vatsal2021} studied the bubble-bursting process in a viscoplastic fluid and revealed how viscoplasticity influences the inertio-capillary waves that drive the bubble-bursting mechanism. Unlike Newtonian liquids, the cavity can sustain shear stress (and a non-flat final shape) if the driving stresses inside the pool fall below the yield stress.

Yield-stress liquids often exhibit elastic behaviour below the yield criterion and also after yielding~\citep{larson1999}. This characteristic gives rise to a distinct subset within the category of yield-stress fluids, referred to as elasto-viscoplastic (EVP) fluids, which behaves akin to an elastic solid below the critical stress identified by the yield stress while exhibiting a viscoelastic fluid behaviour above the yield stress. Different models have been proposed to constitute the behaviour of EVP fluids based on various steady and unsteady flow responses. \hlrev{In the present study, we utilize the EVP model proposed by~\cite{saramito2007}. This model behaves as a Kelvin–Voigt viscoelastic solid prior to yielding and transitions to a \emph{non-linear} viscoelastic liquid in the yielded region, exhibiting Oldroyd-B viscoelastic behaviour far beyond the yield point.} A detailed review of the different EVP models can be found in~\cite{fraggedakis2016}. The physics of elasto-viscoplastic fluids have been explored in a variety of problem sets, such as droplet deformation, deformation in shear flow~\citep{izbassarov2020}, particle migration~\citep{chaparian2020}, channel flow~\citep{izbassarov2021}, bubble migration~\citep{feneuil2023},  porous media flow~\citep{chaparian2020d}, rising bubble~\citep{moschopoulos2021}, and droplet spreading~\citep{oishi2023}. 

Given that a yield-stress fluid exhibits elastic behaviour, it becomes imperative to understand the role of elastic-stress-relaxation in driving capillary wave propagation, which, in turn, influences the formation of jets and droplets, which are the critical characteristics observed in the bubble-bursting process. This study expands the present understanding of bubble bursting (and in general interfacial flows) of elasto-viscoplastic fluids, towards more realistic situations,  which may exhibit additional phenomena such as \emph{shear thinning} and complex initial bubble shape. 

The paper is organised as follows. The methodology and the description of the problem are introduced in~\S\ref{sec:methodology}. In~\S\ref{sec:results}, the results obtained with different combinations of dimensionless elastic stress relaxation time and dimensionless yield stress of the EVP fluid are discussed. The subsections~\S\ref{subsec:vatsal_validation}--\ref{subsec:beta_variation} delve into identifying the influence of the fluid properties on key bursting characteristics. The different modes of energy transfer are presented in~\S\ref{subsec:energy_budget}. The summary and conclusions of the present work are highlighted in~\S\ref{sec:conclusion}. Additional details about the derivation of governing equations, the log-conformation technique to solve for the extra stress in EVP, grid convergence of results, comparison between Bingham and EVP model at the viscoplastic limit, derivation of energy budget terms, energy-based analysis to understand the behaviour of maximum jet height can be respectively found in Appendix~\ref{appendix:non_dimensional_ge}--\ref{appendix:jet_growth}.

\section{Numerical framework}\label{sec:methodology}

\subsection{Conservation laws and constitutive equations}\label{subsec:governing_equations}

A small axisymmetric bubble with an initial radius of $R_0$ is placed on the surface of an incompressible elasto-viscoplastic fluid~(see \S\ref{subsec:initial_condition}). For the considered problem, the dimensionless governing equations are:

\begin{align} \label{eqn:continuity_nonD}
	\boldsymbol{\nabla\cdot u}&=0\,,\\
	\label{eqn:momentum_nonD} \frac{\partial \boldsymbol{u}}{\partial t} + \boldsymbol{\nabla\cdot}\left(\boldsymbol{uu}\right) &= -\boldsymbol{\nabla}p + \boldsymbol{\nabla\cdot}\left(\boldsymbol{\tau_s} + \boldsymbol{\tau_p} \right) + \boldsymbol{f}\,,
\end{align}

\noindent where the velocity field $\boldsymbol{u}$ and the time $t$ are normalized using the inertio-capillary velocity ($V_\gamma = \sqrt{\gamma/\rho_l R_0}$) and time ($T_\gamma = \sqrt{\rho_lR_0^3/\gamma}$) scales, respectively~(here, $\gamma$ and $\rho_l$ are the surface tension and density of the liquid medium, respectively. See Appendix \ref{appendix:non_dimensional_ge} for details). The pressure $p$, the solvent stress $\boldsymbol{\tau_s}$, and the elastic stress $\boldsymbol{\tau_p}$ are normalized using the capillary stress $p_\gamma = \gamma/R_0$. 
\hlrev{Lastly, $\boldsymbol{f} = \boldsymbol{f}_g + \boldsymbol{f}_\gamma$ contains the contributions from gravity $\boldsymbol{f}_g = -\mathcal{B}o\boldsymbol{\hat{e}_z}$, where the Bond number}

\hlrev{ 
\begin{align}
	\mathcal{B}o = \frac{\rho_l g R_0^2}{\gamma}\,,
\end{align}

\noindent with $g$ as acceleration due to gravity, is the ratio between hydrostatic and capillary pressures, and surface tension $\boldsymbol{f}_\gamma = \kappa \delta_s \boldsymbol{\hat{n}}\,$. 
Here, $\kappa$ is the curvature of the liquid-gas interface having a normal vector $\boldsymbol{\hat{n}}$ and $\delta_s$ is a Dirac delta function concentrated at the interface.}

In equation~\eqref{eqn:momentum_nonD}, the deviatoric viscous stress tensor is:

\begin{equation}
	\label{eqn:SolventRheo}
	\boldsymbol{\tau_s} = 2 \text{\Oh}_s \boldsymbol{\mathcal{D}}\,,
\end{equation}

\noindent where $\boldsymbol{\mathcal{D}} = \left(\boldsymbol{\nabla u} + \left(\boldsymbol{\nabla u}\right)^T\right)/2$ is the deformation rate tensor and $\text{\Oh}_s$~denotes the solvent-Ohnesorge number which measures the inertial-capillary timescale compared to the inertial-viscous timescale as defined by

\begin{align}\label{eqn:ohs}
	\text{\Oh}_s = \frac{\mu_s}{\sqrt{\rho_l \gamma R_0}}\,.
\end{align}

\noindent Here, 
$\mu_s$~identifies the solvent-viscosity of the fluid with the total viscosity of the fluid~$\left(\mu_l = \mu_s + \mu_p\right)$ including the contribution from polymeric-viscosity term~$\left(\mu_p\right)$. We can also define a polymeric-Ohnesorge number $\text{\Oh}_p$ given by

\begin{align}\label{eqn:ohp}
	\text{\Oh}_p = \frac{\mu_p}{\sqrt{\rho_l \gamma R_0}}\,,
\end{align}

\noindent based on polymeric viscosity. Consequently, the ratio of solvent to total viscosity is

\begin{align}
	\beta = \frac{\mu_s}{\mu_s + \mu_p} = \frac{\text{\Oh}_s}{\text{\Oh}_s + \text{\Oh}_p} = \frac{\text{\Oh}_s}{\text{\Oh}}\,,
\end{align}

\noindent where $\text{\Oh} = \text{\Oh}_s + \text{\Oh}_p$ corresponds to the Ohnesorge number based on the total viscosity of the fluid.

The extra stress tensor $\boldsymbol{\tau_p}$ embeds the elastic and plastic behaviour of the EVP fluid and is modelled with the constitutive relationship proposed by~\cite{saramito2007}. Using an order parameter $\boldsymbol{A}$ (conformation tensor) tracking the stretch of the EVP matrix~\citep{snoeijer2020, Stone2023Note} with a base state $\boldsymbol{A} = \boldsymbol{I}$ (here, $\boldsymbol{I}$ is the identity tensor), this constitutive model can be summarised as

\begin{align}
	\label{eqn:Aconform}
	\boldsymbol{\overset{\nabla}{A}} &= -\frac{K}{\text{\De}}\left(\boldsymbol{A} - \boldsymbol{I}\right),\,\,\text{with}\,K = \mathrm{max}\left(\frac{\|\boldsymbol{\tau_d}\|-\mathcal{J}}{\|\boldsymbol{\tau_d}\|},0\right),\,\text{and}\\
	\label{eqn:Stress_Conform}
	\boldsymbol{\tau}_p &= \frac{\text{\Oh}_p}{\text{\De}}\left(\boldsymbol{A} - \boldsymbol{I}\right)\,.
\end{align}

\noindent Here, the Deborah number

\begin{align}
	\text{\De} = \frac{\lambda}{T_\gamma}
\end{align}

\noindent is the dimensionless relaxation time $\lambda$ of the EVP matrix to its base state $\boldsymbol{A} = \boldsymbol{I}$, normalized using the inertial-capillary timescale $T_\gamma$. In equation~\eqref{eqn:Aconform}, $\boldsymbol{\overset{\nabla}{A}}$~is the upper-convected time derivative and $\|\boldsymbol{\tau_d}\|$~is the second invariant of the deviatoric part of the elastic stress tensor and are defined as,

\begin{align}
	\label{eqn:upper_convected_derivative}
	\boldsymbol{\overset{\nabla}{A}} &= \frac{\partial \boldsymbol{A}}{\partial t} + \left(\boldsymbol{u\cdot\nabla}\right)\boldsymbol{A} - \boldsymbol{A\cdot}\left(\boldsymbol{\nabla u}\right) - \left(\boldsymbol{\nabla u}\right)^T\boldsymbol{\cdot}\boldsymbol{A}\,\text{and}\\
	\|\boldsymbol{\tau_d}\| &= \sqrt{\frac{\boldsymbol{\tau_d} : \boldsymbol{\tau_d}}{2}}\,,
\end{align}

\noindent respectively. \hlrev{Here, $\boldsymbol{\nabla u} := \partial u_j/x_i$ in Einstein notation.} The deviatoric part of the elastic stress tensor is calculated as $\boldsymbol{\tau_d} = \boldsymbol{\tau_p} - \left(\mathrm{tr} \left(\boldsymbol{\tau_p}\right)/\mathrm{tr}\left(\boldsymbol{I}\right)\right)\boldsymbol{I}$. Lastly, the plastocapillary number~$\left(\mathcal{J}\right)$ accounts for the competition between the yield stress $\tau_y$ and the Laplace pressure $\gamma/R_0$ as

\begin{align}
	\mathcal{J} &=\frac{\tau_y R_0}{\gamma}.
\end{align}

\noindent In equation~\eqref{eqn:Aconform}, $K$ is a dimensionless function that acts as a stress-dependent switch, controlling the transition from viscoelastic solid-like to viscoelastic fluid-like behaviour in the EVP fluid. Consequently, in the yielded state $\text{\De}/K$ can be interpreted as the effective relaxation time. Consequently, below the yield stress ($K = 0$), $\boldsymbol{\overset{\nabla}{A}} = \bold{0}$ and the EVP matrix deform according to the flow field \citep[see equation~\eqref{eqn:upper_convected_derivative} and][]{Stone2023Note} but do not relax. Additionally, the stress $\boldsymbol{\tau_p} = \text{\Oh}_p(\boldsymbol{A} - \boldsymbol{I})/\text{\De}$ depends only on the elastic (polymeric) deformation and the elasto-capillary number

\begin{align} \label{eqn:elasto_capillary_number}
	Ec = \frac{GR_0}{\gamma} = \frac{\text{\Oh}_p}{\text{\De}}\,,
\end{align}

\noindent where $G = \mu_p/\lambda$ is the elastic modulus. Above the yield stress ($K > 0$), the EVP fluid behaves like a viscoelastic liquid following the constitutive relation given by combining equations~\eqref{eqn:Aconform}--\eqref{eqn:Stress_Conform} to give

\begin{align} \label{eqn:polymer_stress_evolution}
	\text{\De}\, \boldsymbol{\overset{\nabla}{\tau}_p} + K\boldsymbol{\tau_p} = 2\text{\Oh}_p\, \boldsymbol{\mathcal{D}}\,.
\end{align}

We also model the gas-phase with the corresponding conservation laws, which are similar to equations~\eqref{eqn:continuity_nonD} and~\eqref{eqn:momentum_nonD}~(see Appendix~\ref{appendix:non_dimensional_ge}). We keep the gas-liquid density ratio $\rho_r$~$\left(=\rho_g/\rho_l\right)$ fixed at $10^{-3}$. Similarly, the viscosity ratio $\mu_r$~$\left(=\mu_g/\mu_l\right)$ is set constant at $2\times 10^{-2}$ throughout the work. 










\subsection{Simulation setup} \label{subsec: sim_setup}

The direct numerical simulations are performed with the open-source free-software \hlrev{language} Basilisk C~\citep{popinet2009, popinet2013}, which offers adaptive mesh refinement (AMR) based on wavelet estimated discretization errors, making it well-suited for singular interfacial flows~\citep{berny2020,VatsalThesis,yang2023enhanced}. 
Basilisk C uses a volume of fluid (VoF) technique to track the interface with the help of a colour function $c$ ($c = 1$ in liquid and $c = 0$ in gas), which satisfies the scalar-advection equation. The geometrical features of the interface such as its unit vector normal $\boldsymbol{\hat{n}}$ and the curvature $\kappa$ ($= \boldsymbol{\nabla\cdot\hat{n}}$) are calculated using the height-function method \citep{popinet2009, popinet2018}. The governing equations for the gas and the fluid are solved using a one-fluid approximation \citep{prosperetti2009computational, tryggvason2011}, \hlrev{where the singular surface tension is approximated as $\boldsymbol{f}_\gamma = \kappa\delta_s\boldsymbol{\hat{n}} \approx \kappa\boldsymbol{\nabla}c$ \citep{brackbill1992}}. Note that the time step in our simulations is restricted by the oscillation period of the smallest wavelength of the capillary wave because the surface tension scheme is explicit in time~\citep{popinet2009,popinet2013}. 

Utilizing the AMR feature of Basilisk C, the errors in the VoF tracer and interface curvature are minimized by applying a tolerance threshold of $10^{-3}$ and $10^{-4}$, respectively. In addition, the refinement of the grid is also performed based on the velocity (tolerance threshold: $10^{-2}$), conformation tensor~$\boldsymbol{A}$ (tolerance threshold: $10^{-2}$) and yielded region identified by \hlrev{$K$} (tolerance threshold: $10^{-3}$) to accurately resolve the regions of low strain rates and elastic deformation. \hlrev{These tolerance threshold values can be interpreted as the maximum error associated with the subsequent application of volume-averaged downsampling of fine-resolution-solution and bi-linear upscaling of coarse-level-solution \citep{popinet2015, van2018towards}. We highlight that these refinements offer the advantage of an almost uniform grid in key regions of interest~(see also~\cite{antoon}) and acknowledge that the efficacy of such refinement criteria as employed in this study needs further investigation.} For AMR, we employ a grid resolution ensuring a minimum cell size of $\Delta = R_0/1024$, corresponding to 1024 cells across the initial bubble radius. However, when $\text{\De} \ge 1$, we switch to $\Delta = R_0/2048$. Comprehensive grid-independence studies were conducted to confirm that the results remain unaffected by the chosen grid size (see Appendix~\ref{appendix:grid_independence}). We consider a square domain measuring $8R_0$ on each side, representing only one slice of the three-dimensional bursting bubble process leveraging the axisymmetric flow assumption. For both liquid and gas, free-slip and no-penetration boundary conditions are applied at the domain boundaries, while a zero-gradient condition is used for pressure. To ensure that ejected droplets, which arise from the breakup of the Worthington jet, can leave the domain, an outflow boundary condition is employed at the top boundary. The chosen domain size ensures that the boundaries do not influence the bubble bursting process. 
\hlrev{Lastly, the solution of the constitutive relations (equation~\eqref{eqn:Aconform}-\eqref{eqn:Stress_Conform}) require the log-conformation approach proposed by \citet{fattal2004}~(also see \citet{lopez2019,oishi2023,dixit2024}; and appendix~\ref{appendix:log_conformation}). For details of our implementation in Basilisk C, we refer the readers to~\citet{ariGithub}.}

\subsection{Initial condition} \label{subsec:initial_condition}

The initial shape of the bubble is obtained by solving the Young--Laplace equations to find the quasi-static equilibrium state for a specified Bond number, $\mathcal{B}o$~(see \citet{villermaux2012, deike2018, vatsal2021}). In this study, we focus more on the capillary effects than gravitational effects, and hence the value of $\mathcal{B}$o is set to $10^{-3}$. At low values of $\mathcal{B}$o, the initial shape of the bubble closely approximates a sphere within the surrounding Newtonian medium. However, for the elasto-viscoplastic medium considered in this study, we make a significant assumption by retaining the spherical bubble shape. It is important to note that this assumption is a crucial aspect of our investigation. Given the elasto-viscoplastic behaviour of the liquid medium in our work, the bubble shape close to the fluid-air interface would exhibit a different interface profile than that in the Newtonian medium. The shapes of the bubble rising in an EVP medium constitute an active area of research~(see \cite{lopez2018,daulet2018,moschopoulos2021}). 

Further, the bubble's shape at the fluid-air interface depends not only on the material behaviour of the surrounding medium in which the bubble rises but also on the generation and dynamics before reaching the free surface. The bubble rise depends on the buoyancy forces, which should be strong enough to yield the elastoviscoplastic medium. Hence, one might expect a non-trivial shape as suggested by~\cite{lopez2018,deoclecio2023drop}. However, in the present study, we consider the small spherical bubbles trapped at the interface rather than rising bubbles reaching the free surface to be consistent with the previous investigations. It should be pointed out that the bubble cap breakup is also sensitive to the employed numerical method. For a comparative analysis to understand the transient effects of elasticity on the bubble bursting in an elasto-viscoplastic medium, we consider the same initial condition as employed by~\cite{vatsal2021} for the case of bubble bursting in a viscoplastic medium. Hence, an initial stress-free condition is employed in the present computations, which would otherwise correspond to some pre-sheared state if the bubble rose close to the free surface and burst instantly.

The bubble-bursting problem could also be coupled with the bubble rise in an elastoviscoplastic medium with a free surface to obtain a more realistic initial condition. However, such a coupling is limited by the numerical methodology to treat the breakup of bubble cap~$\delta$~(as the thin initial film between the bubble and air drains). Moreover, the bubbles usually sit on the free surface before the cap bursts \citep{bartlett2023universal}, allowing enough time for the elastic stresses to relax if the drainage time is much longer than the relaxation time. Hence, as a simplification in this work, we consider the open cavity the initial condition, as shown in figure~\ref{fig_initial_condition}. In this figure, $\left(\mathcal{R},\mathcal{Z}\right)$ represent the radial and axial coordinate system, and $\mathcal{H}_i \approx 2$ denotes the initial bubble depth, while $\theta_i$ indicates the initial location of the cavity-free surface intersection. We incorporate a finite curvature~$\kappa_0=100$ at the intersection of bubble and free surface to regularize the curvature singularity, consistent with~\cite{vatsal2021}, which has been demonstrated to have no significant influence on the bubble-bursting dynamics.

\section{Results} \label{sec:results}

\subsection{Validation} \label{subsec:vatsal_validation}
\begin{figure}
\centering
\resizebox*{0.75\linewidth}{!}{\includegraphics{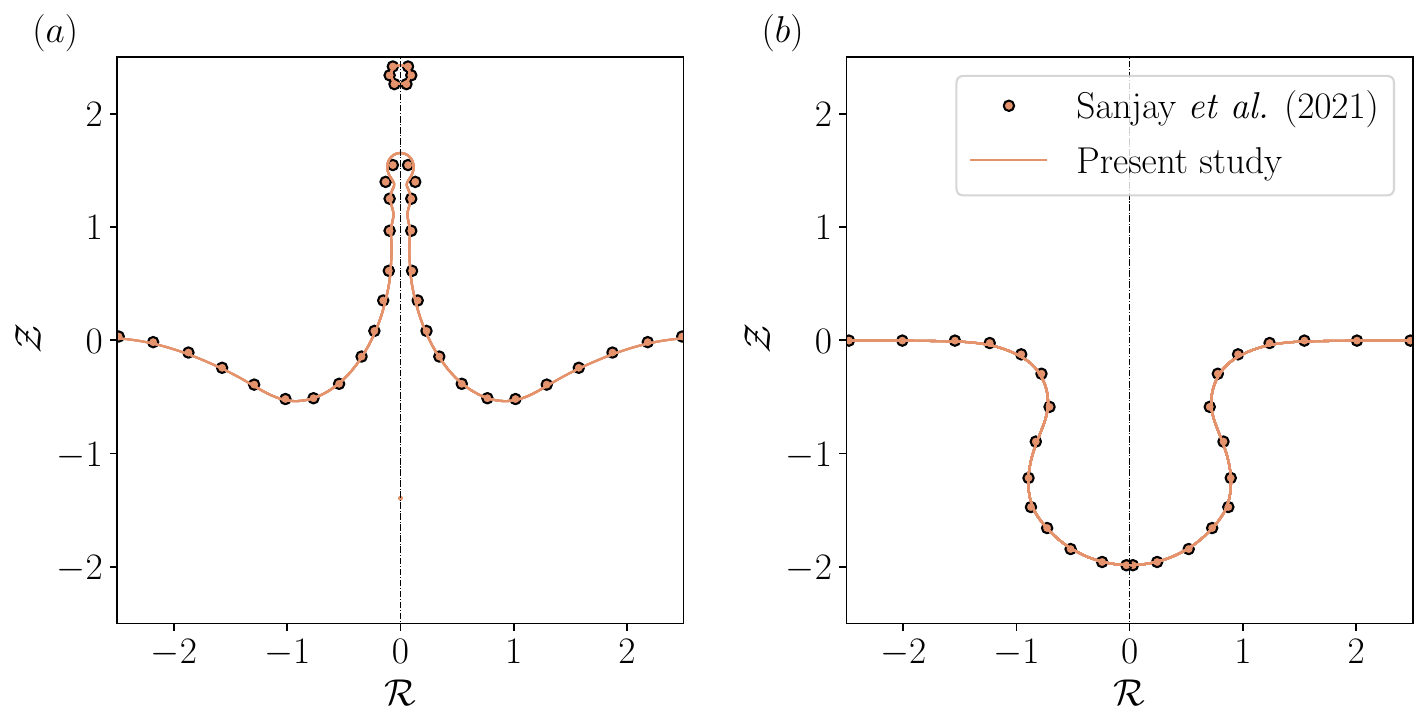}}
\caption[]{Validation of interface shapes obtained with elasto-viscoplastic fluid at very-low Deborah number of $\text{\De}=10^{-4}$ against the results obtained by~\cite{vatsal2021} with viscoplastic fluid at $(a)$ $t=1.0,\;\mathcal{J}=0.1$, $(b)$ $t=0.75,\;\mathcal{J}=1.0$.
} 
\label{fig_vatsal_validation}
\end{figure}

\begin{figure}
\centering
\resizebox*{0.95\linewidth}{!}{\includegraphics{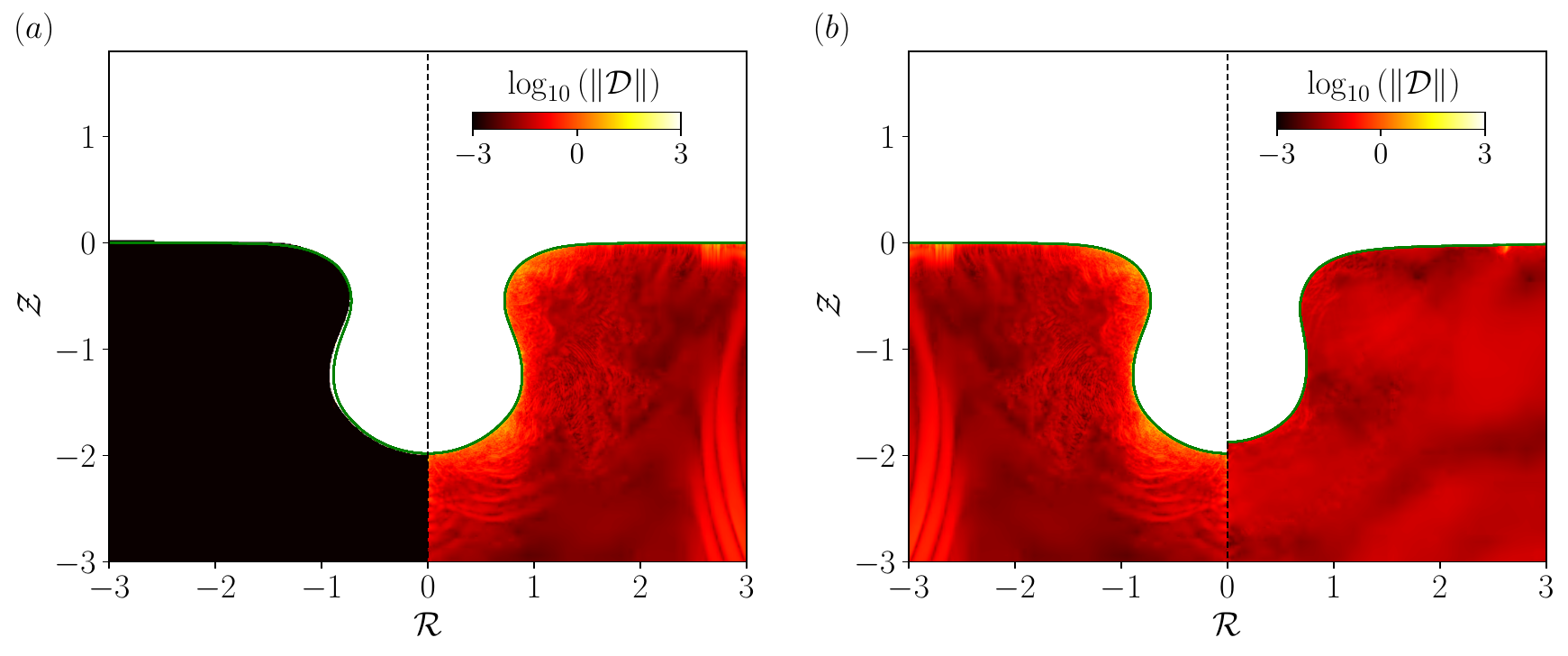}}
\caption[]{Comparison of the deformation-rate tensor obtained with (a, left panel) viscoplastic fluid by~\cite{vatsal2021} and (a, right panel) elasto-viscoplastic fluid at $\mathcal{J} = 1.0,\,\text{\De}=10^{-4}$ at $t=1.0$. Time evolution of deformation tensor in EVP fluid at (b, left panel)~$t=1.0$ and (b, right panel)~$t=1.6$ for $\mathcal{J} = 1.0,\,\text{\De}=10^{-4}$.
} 
\label{fig_vatsal_validation_D2}
\end{figure}

This section compares our results with \citet{vatsal2021}, who used a regularized Bingham model to study bubble bursting in a viscoplastic medium. In such a viscoplastic model, the fluid features a rigid body motion ($\boldsymbol{\mathcal{D}} = \bold{0}$) below the yield stress and flows like a viscous liquid above the yield stress. The EVP model used in this work (equations~\eqref{eqn:SolventRheo} and~\eqref{eqn:polymer_stress_evolution}) reduces to the Bingham model if $\text{\De} = 0$ and $\text{\Oh}_s = 0$~(see also Appendix~\ref{appendix:Bingham_EVP}), giving 
	
\begin{equation} \label{eqn:apparent_viscosity_J}
	\boldsymbol{\tau} =	\boldsymbol{\tau_p} + 	\boldsymbol{\tau_s} = 2\left(\text{\Oh}_s +  \frac{\text{\Oh}_p}{K}\right)\boldsymbol{\mathcal{D}} = 2\frac{\text{\Oh}_p}{K}\boldsymbol{\mathcal{D}}\,,
\end{equation}
	
\noindent where $\left(\text{\Oh}_s + \text{\Oh}_p/K\right) = \text{\Oh}_p/K$ is the apparent viscosity. Figures~\ref{fig_vatsal_validation} and~\ref{fig_vatsal_validation_D2} illustrate the comparison between \citet{vatsal2021} and our simulations with $\text{\De}=10^{-4}$, $\text{\Oh}_p = 9.5 \times 10^{-3}$, and $\text{\Oh}_s=5\times 10^{-4}$. For a low plastocapillary number ($\mathcal{J} = 0.1$), the capillary waves meet at the bottom of the bubble cavity, leading to an inertial flow-focusing that forms a Worthington jet, which subsequently breaks into droplets (figure~\ref{fig_vatsal_validation}$a$). 

On the other hand, at high $\mathcal{J}$ (\textit{i.e.} at $\mathcal{J} = 1$), at comparable timescales between the two models, the bubble cavities are identical (figures~\ref{fig_vatsal_validation}$b$ and~\ref{fig_vatsal_validation_D2}$a$). Nonetheless, the two cases show different $\|\boldsymbol{\mathcal{D}}\|$. This apparent discrepancy can be attributed to the different behaviours of the unyielded region in the Bingham viscoplastic and the \cite{saramito2007} elasto-viscoplastic models. Notably, in the case of the regularized Bingham model employed for simulating bubble bursting in viscoplastic fluid by~\cite{vatsal2021}, at the stoppage time, the entire medium is unyielded ($K \approx 0$ throughout the bulk) and the fluid flow ceases due to stresses falling below the yield stress. Consequently, the deformation tensor~$\|\boldsymbol{\mathcal{D}}\|$ is zero (rigid body rotation \hlrev{or no flow}). 
For the elasto-viscoplastic model, nearly the entire liquid remains unyielded as regions with $K \neq 0$ are predominantly situated close to the fluid interface. However, the bulk exhibits a Kelvin--Voigt viscoelastic solid behaviour \citep[$\boldsymbol{\overset{\nabla}{A}} = \bold{0}$  and $\boldsymbol{\tau_p} = Ec(\boldsymbol{A} - \boldsymbol{I})$, see][]{saramito2007}. Consequently, even below the yield stress, the deformation tensor~$\|\boldsymbol{\mathcal{D}}\|$ can be non-zero. Note that, for the Kelvin--Voigt viscoelastic rigid body motion ($\|\boldsymbol{\mathcal{D}}\| = 0$) occurs in the limit of very large elastic modulus (i.e., $Ec = \text{\Oh}_p/\text{\De} \to \infty$). \hlrev{The simulations are stopped at $t=2$ and the final interface shape of the bubble, where the extra stress balances the capillary stress, are not investigated in this study.}

\subsection{Regime map} \label{subsec:regime_map}

\begin{figure}
	\centering
	\resizebox*{1.0\linewidth}{!}{\includegraphics{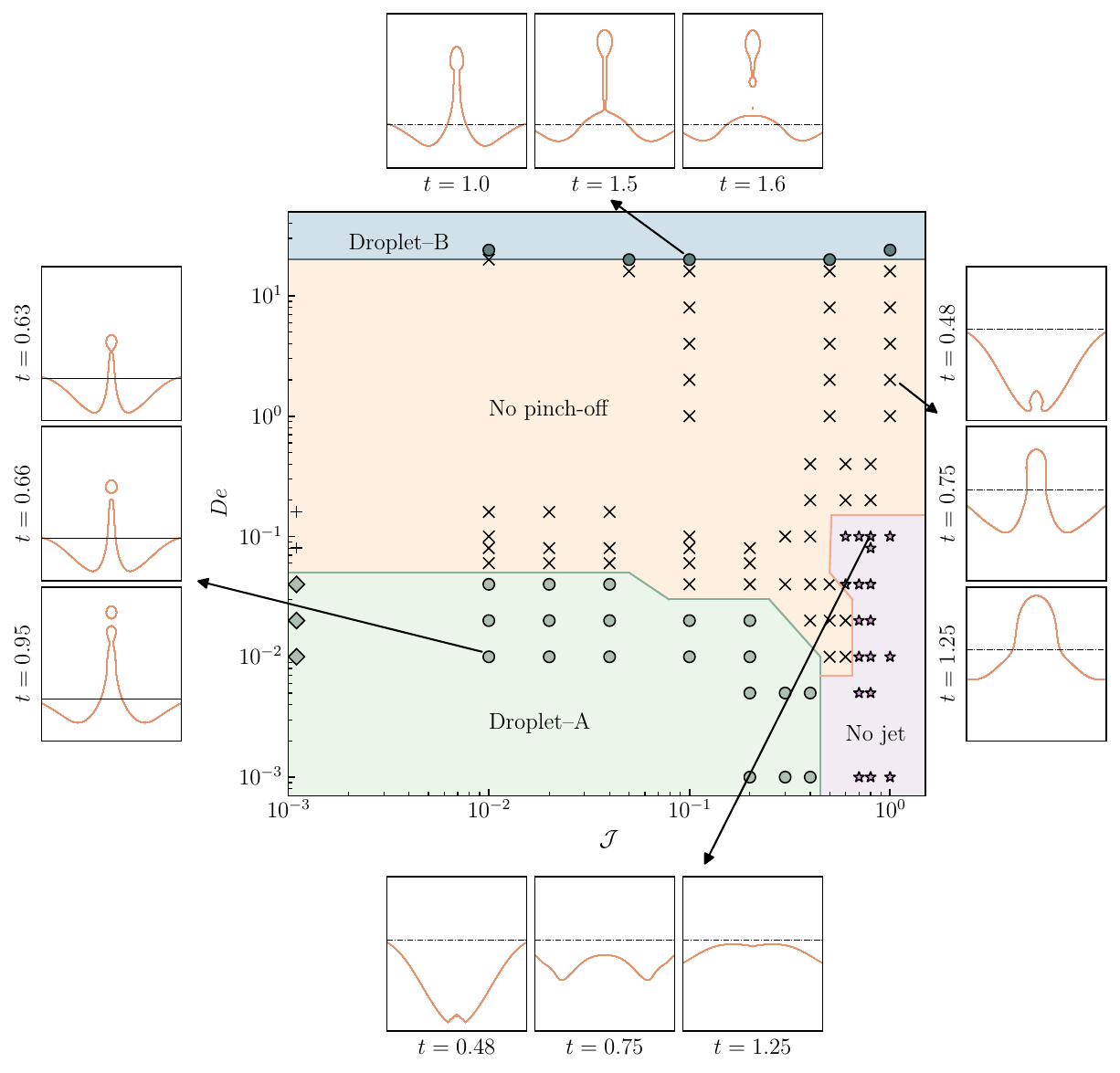}}
	\caption[]{
		Regime map in terms of the Plastocapillarity number $\mathcal{J}$ and the Deborah number $\text{\De}$, distinguishing the droplet formation (Droplet--I), no jet formation, jet pinch-off (Droplet--II) and no-pinch-off regimes. All the cases correspond to $\text{\Oh}_s = \text{\Oh}_p = 0.005$. The four series of insets illustrate typical cases in these regimes, namely ($\mathcal{J}$, $\text{\De}$) = (0.1, 20) for the upper, (0.01, 0.01) for the left, (0.8, 0.1) for the bottom and (1, 2) for the right series of images, respectively. Different markers identify the viscoelastic limit corresponding to~$\mathcal{J}=0$.} 
	\label{fig_regime_map}
\end{figure}

We investigate the dynamics of bubble bursting in an elasto-viscoplastic medium by exploring the influence of elastic stress relaxation and yield stress, quantified by the Deborah~$\text{\De}$ and the plastocapillary~$\mathcal{J}$ numbers, respectively \citep[see][]{oishi2023}. Our exploration spans a parameter space with~$\text{\De}\in\left[10^{-3},20\right]$ and~$\mathcal{J} \in \left[10^{-2},1\right]$ while maintaining a fixed Ohnesorge number of~$\text{\Oh}=10^{-2}$, $\beta$ of 0.5, and Bond number of~$\mathcal{B}o=10^{-3}$. The value of ~$\text{\Oh}_s = \text{\Oh}_p = 0.005$ (which fixes $\beta$ at 0.5) is applicable for all the discussions except for~$\S$\ref{subsec:beta_variation} \hlrev{(see ~$\S$\ref{subsec:beta_variation} to identify the variation of regime map with~$\beta$)}. All the simulations were carried out until~$t\ge 1.2$, as this time was seen to be sufficient to capture the key dynamics of the bursting process. This investigation leads us to a regime map, presented in figure~\ref{fig_regime_map}. We identify four distinct regimes, namely—(i) \emph{droplet} formation from the tip of the Worthington jet (Droplet--A), (ii) \emph{no Worthington jet} formation, (iii) \emph{no pinch-off} of the Worthington jet and no droplet, (iv) \emph{pinch-off} at the base of the Worthington jet to form a droplet (Droplet--B). Note that in the context of this study, we characterize a jet formed by inertial flow-focusing as the \emph{Worthington jet} if it crosses the equilibrium surface $\mathcal{Z}=0$ at the axis~$(\mathcal{R}=0)$.

\subsubsection{Droplet formation regime, Droplet--A (Low elastic stress relaxation time and yield-stress limit, $\mathcal{J} \rightarrow 0,\,\text{\De}\rightarrow 0$)}
\label{subsec:RegimeI}
For small Deborah and plastocapillary numbers ($\text{\De} \le 0.05$ and $\mathcal{J} \le 0.4$), we observe a Newtonian-like behaviour where the initial capillary waves collapse at the bottom of the bubble cavity resulting in a Worthington jet formation \citep{Gordillo2023,dixit2024}. Further, due to the higher capillary forces compared to the viscous and elastic forces, the jet breaks up, 
resulting in droplets~\citep{walls2015}. This regime often features multiple drops similar to the case of the Newtonian bubble bursting process \citep{berny2020}. 

For the viscoplastic fluid,~\citet{vatsal2021} had identified this regime to fall below~$\mathcal{J} \approx 0.3$ for small $\text{\Oh}$ and $\text{\De} = 0$, beyond which the jet breakup is suppressed owing to an increase in the apparent viscosity which critically dampens the capillary waves. Figure~\ref{fig_regime_map} identifies this transition at $\mathcal{J} \approx 0.5$ for $\text{\De} \to 0$. This delayed droplet--no-droplet transition is attributed to a reduction in the effective viscosity of the EVP matrix in comparison to the purely Bingham fluid~(refer Appendix~\ref{appendix:Bingham_EVP}). The effective viscosity in the limit of vanishing $\text{\De}$ is $\text{\Oh}_p/K$ contrary to $\text{\Oh}/K$ for purely Bingham fluid \citet{vatsal2021}. This delay in stress relaxation delays the transition to the plastic behaviour within the EVP matrix. Consequently, at finite $\text{\De}$, a critical increase in the apparent viscosity occurs at a higher value of $\mathcal{J}$.

Furthermore, as $\text{\De}$ increases further, even for low $\mathcal{J}$, jet breakup into droplets is suppressed. This observation agrees with that of~\citet{rodriguez2023} and~\citet{dixit2024}: adding polymers hinders droplet ejection even for the small solvent-to-polymer viscosity ratio. We attribute this observation to a delay in elastic stress relaxation at higher $\text{\De}$, increasing the elastic stresses that counteract capillarity to prevent both the end-pinching and Rayleigh-Plateau instabilities \citep{PandeySM2021}.

\subsubsection{No-jet regime (Viscoplastic limit, $\mathcal{J} \gg 0,\,\text{\De}\rightarrow 0$)}

In the case of a purely viscoplastic fluid, \citet{vatsal2021} observed that for~$\mathcal{J} \ge 0.65$, the surface tension fails to yield the entire cavity, and the capillary wave freezes before reaching the bottom of the cavity, leading to a non-flat final equilibrium shape. In this work, the \emph{no jet} regime commences at $\mathcal{J} \approx 0.7$ in the limit of $\text{\De} \to 0$, characterized by the absence of a jet crossing the free surface ($\mathcal{Z} = 0$). This finding aligns with the increased plasticity effect at higher $\mathcal{J}$.

However, our results diverge from the case of a purely viscoplastic fluid in one critical aspect: we find that despite the increasing plasticity, the bubble cavity consistently yields, albeit slowly. In the context of purely viscoplastic fluids, the yield surface is stationary and aligns with the cavity boundary, resulting in zero deformation rate ($\|\boldsymbol{\mathcal{D}}\| = 0$), independent of resultant stress field that are below the yield stress. However, under elasto-viscoplastic conditions, the medium behaves akin to a Kelvin--Voigt solid below the yield stress, leading to a slow deformation of the cavity over extended timescales depending on the elastocapillary number $Ec = \text{\Oh}_p/\text{\De} = (1-\beta) \text{\Oh}/\text{\De}$, as indicated by a non-zero deformation rate~($\|\boldsymbol{\mathcal{D}}\| \ne 0$, see~\S\ref{subsec:vatsal_validation}). For an infinite $Ec$, a rigid body motion of the unyielded region is recovered.

\subsubsection{No pinch-off regime (Visco-elasto-capillary limit, $\text{\De} \sim \mathcal{O}(1)$)}

At large values of $\mathcal{J}$, as we increase Deborah number ($\text{\De} \sim \mathcal{O}(1)$), we notice Worthington jet forms irrespective of $\mathcal{J}$. This $\mathcal{J}$--independent behaviour is due to a decrease in the elastocapillary number $Ec = \text{\Oh}_p/\text{\De} = (1-\beta) \text{\Oh}/\text{\De}$ with an increase in $\text{\De}$ at fixed $\text{\Oh}_p$. \hlrev{Consequently, for a considered deformation of the EVP matrix the maximum elastic energy decreases with~$\text{\De}$ at fixed~$\text{\Oh}_p$~(refer figure~\ref{app_fig:visco_elasto_capillary_1}$c$), leading to the formation of the jet.} This jet development is significantly influenced by the variations in the ratio of polymer to total viscosity~($\beta$, see~\S\ref{subsec:beta_variation}), which modifies the elastocapillary number at fixed $\text{\Oh}$ and $\text{\De}$. Nonetheless, the elastic forces still dominate over capillary, resulting in the prevention of droplet formation from the jet even at small $\mathcal{J}$ (see~\S\ref{subsec:RegimeI}). The role of elastocapillary number in jet formation is further explained in~\ref{subsec:bursting_dynamics}.

\subsubsection{Droplet formation regime, Droplet--B (Newtonian-like limit, $\text{\De}\gg 1$)}\label{sec:NewtLike}

For very high values of Deborah numbers ($\text{\De} \gg 1$), the bursting bubble dynamics appear to be independent of $\mathcal{J}$. However, in contrast to the ``\emph{no pinch-off} of the Worthington jet and no droplet regime", the Worthington jet breaks up at the base to form one droplet. In this regime, the yield surface is still very close to the liquid-gas interface~\hlrev{(refer figure~\ref{fig_index_plot_De_1}$d$,~\ref{fig_index_plot_De_2}$c$)}, and most of the EVP fluid remains unyielded. At such high values of $\text{\De}$, the elastocapillary number becomes too small, and the elastic stresses fail to prevent either the formation of the Worthington jet or its subsequent breakup at the base. Such a Newtonian-like regime with vanishing elastic stresses was also found by~\citet{oishi2023}.

\subsection{Bubble-bursting dynamics} \label{subsec:bursting_dynamics}

\begin{figure}
\centering
\resizebox*{1.0\linewidth}{!}{\includegraphics{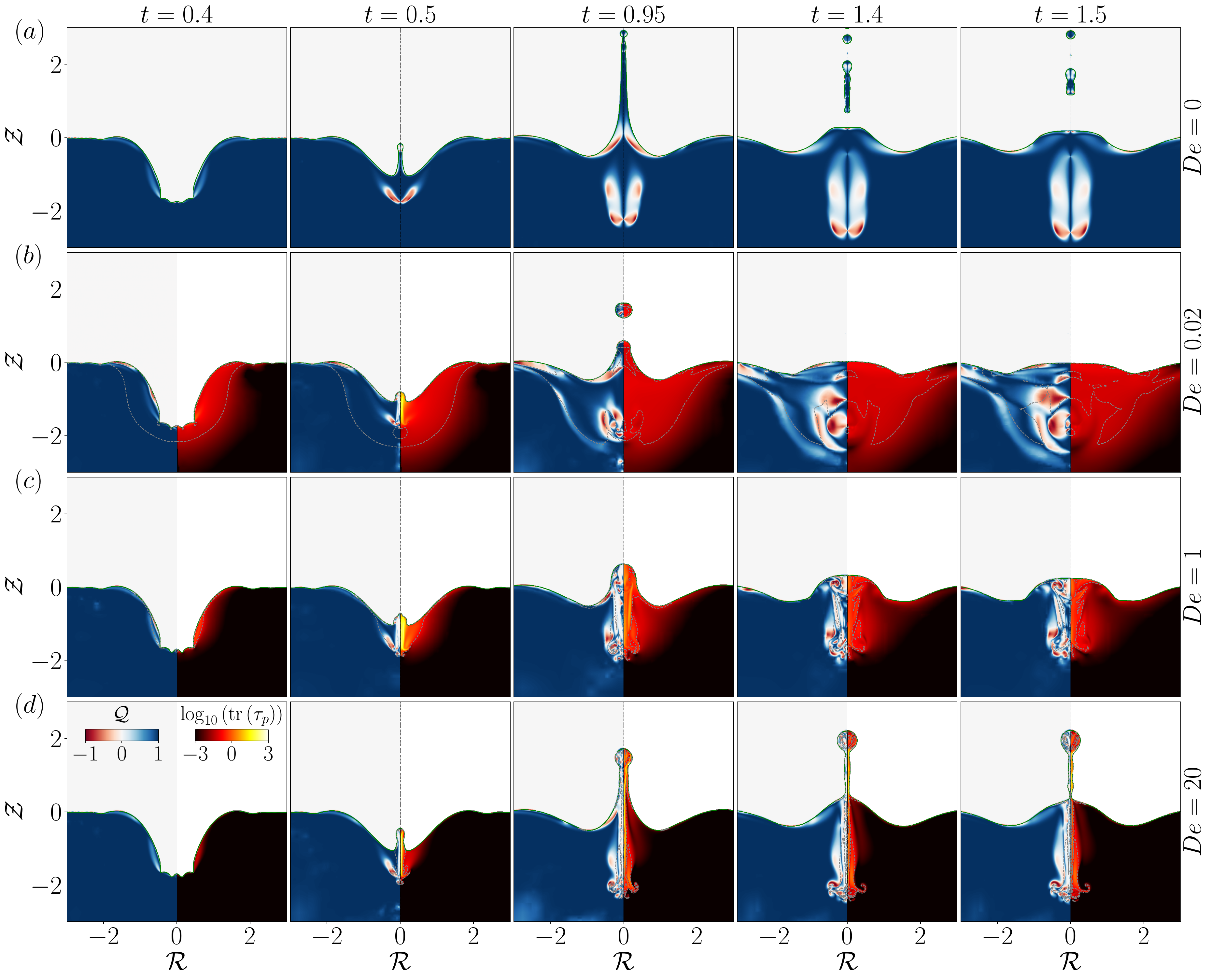}}
\caption[]{Variation of the dynamics of bubble bursting in a $(a)$~Newtonian medium compared against elasto-viscoplastic medium for $\mathcal{J}=0.1$, $\text{\Oh}=10^{-2}$ at $(b)$~$\text{\De}=0.02$, $(c)$~$\text{\De}=1$, $(d)$~$\text{\De}=20$. The left part of each panel shows the flow topology parameter $\mathcal{Q}$ and the right part of the panel shows the trace of elastic stress on a $\mathrm{log_{10}}$ scale. The yielded regions are marked by a grey line, which corresponds to~$\mathcal{J}\approxeq \|\boldsymbol{\tau_d}\|$. For Newtonian medium, both panels show the flow topology parameter.} 
\label{fig_index_plot_De_1}
\end{figure}

In the previous section, we identified different regimes as a function of $\text{\De}$ and~$\mathcal{J}$. Here, we analyze the transient development of bubble bursting process. To demystify the different stages of the bubble bursting process, we will compare the dynamics of bubble bursting in an elasto-viscoplastic medium with that of a Newtonian fluid. The later is well documented in~\cite{duchemin2002,ghabache2016,deike2018,gordillo2019,berny2020}. The bubble bursting in a Newtonian liquid~($\mathcal{J}$ = 0, $\text{\De}$ = 0) is characterised by the retraction of the rim leading to the formation of capillary waves. The capillary waves travel towards the bottom of the bubble cavity resulting in the formation of a Worthington jet, which can then break into multiple droplets owing to the end-pinching and Rayleigh-Plateau instabilities~\citep{walls2015}. Furthermore, owing to the conservation of momentum, a high-velocity jet is also formed in an opposite direction to the Worthington jet, inside the liquid pool. Figure~\ref{fig_index_plot_De_1}$a$ illustrates the process of bubble bursting in a Newtonian fluid medium and identifies the state of flow inside the liquid medium using theflow topology parameter~$\mathcal{Q}$ defined as
\begin{align} \label{eqn:topology_parameter}
    \mathcal{Q} = \frac{\|\mathcal{D}\|^2 - \|\mathcal{S}\|^2}{\|\mathcal{D}\|^2 + \|\mathcal{S}\|^2}\,,
\end{align}
\noindent where~$\|\mathcal{D}\| = \sqrt{(\boldsymbol{\mathcal{D}} \colon \boldsymbol{\mathcal{D}})/2}$ and $\|\mathcal{S}\| = \sqrt{(\boldsymbol{\mathcal{S}:\mathcal{S}})/2}$, with $\boldsymbol{\mathcal{S}}$ denoting the rate of rotation tensor defined as~$\boldsymbol{\mathcal{S}} = \left(\left(\boldsymbol{\nabla u}\right)^T - \boldsymbol{\nabla u}\right)/2$. When~$\mathcal{Q} = -1$, the flow is purely rotational, whereas regions with~$\mathcal{Q}=0$ represent pure shear flow and~$\mathcal{Q} = 1$ corresponds to either elongational flow or no flow~(i.e. $\|\mathcal{D}\| \rightarrow 0, \|\mathcal{S}\| \rightarrow 0$).

\begin{figure}
\centering
\resizebox*{1.0\linewidth}{!}{\includegraphics{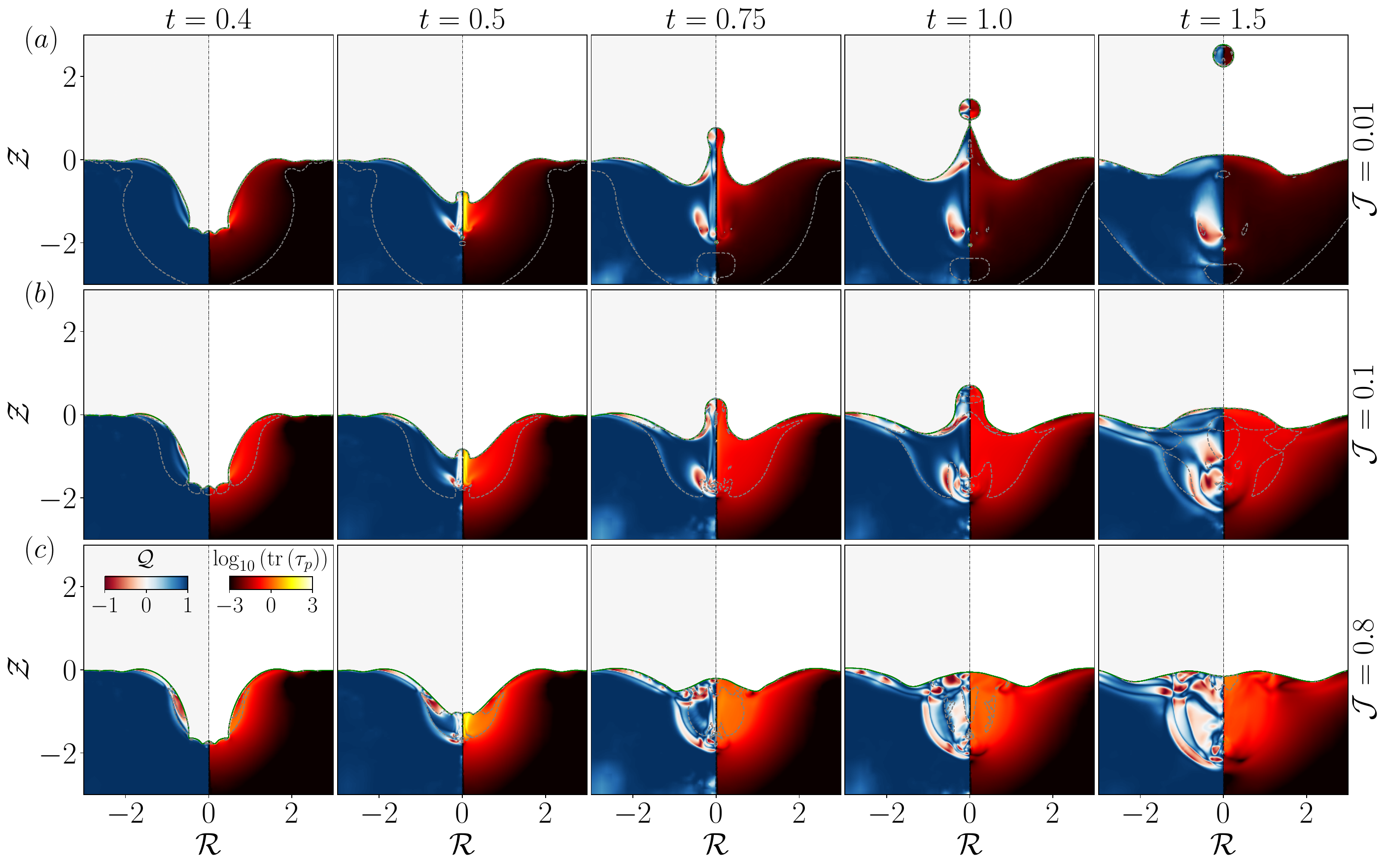}}
\caption[]{Variation of the dynamics of bubble bursting in an elasto-viscoplastic medium with respect to $\mathcal{J}$ at $\text{\De}=0.04$. $(a)$~$\mathcal{J}=0.01$, $(b)$~$\mathcal{J}=0.1$, $(c)$~$\mathcal{J}=0.8$. The left part of each panel shows the flow topology parameter $\mathcal{Q}$ and the right part of the panel shows the trace of elastic stress on a $\mathrm{log_{10}}$ scale. The yielded regions are marked by a grey line, which corresponds to~$\mathcal{J}\approxeq \|\boldsymbol{\tau_d}\|$.} 
\label{fig_index_plot_J}
\end{figure}

\begin{figure}
\centering
\resizebox*{1.0\linewidth}{!}{\includegraphics{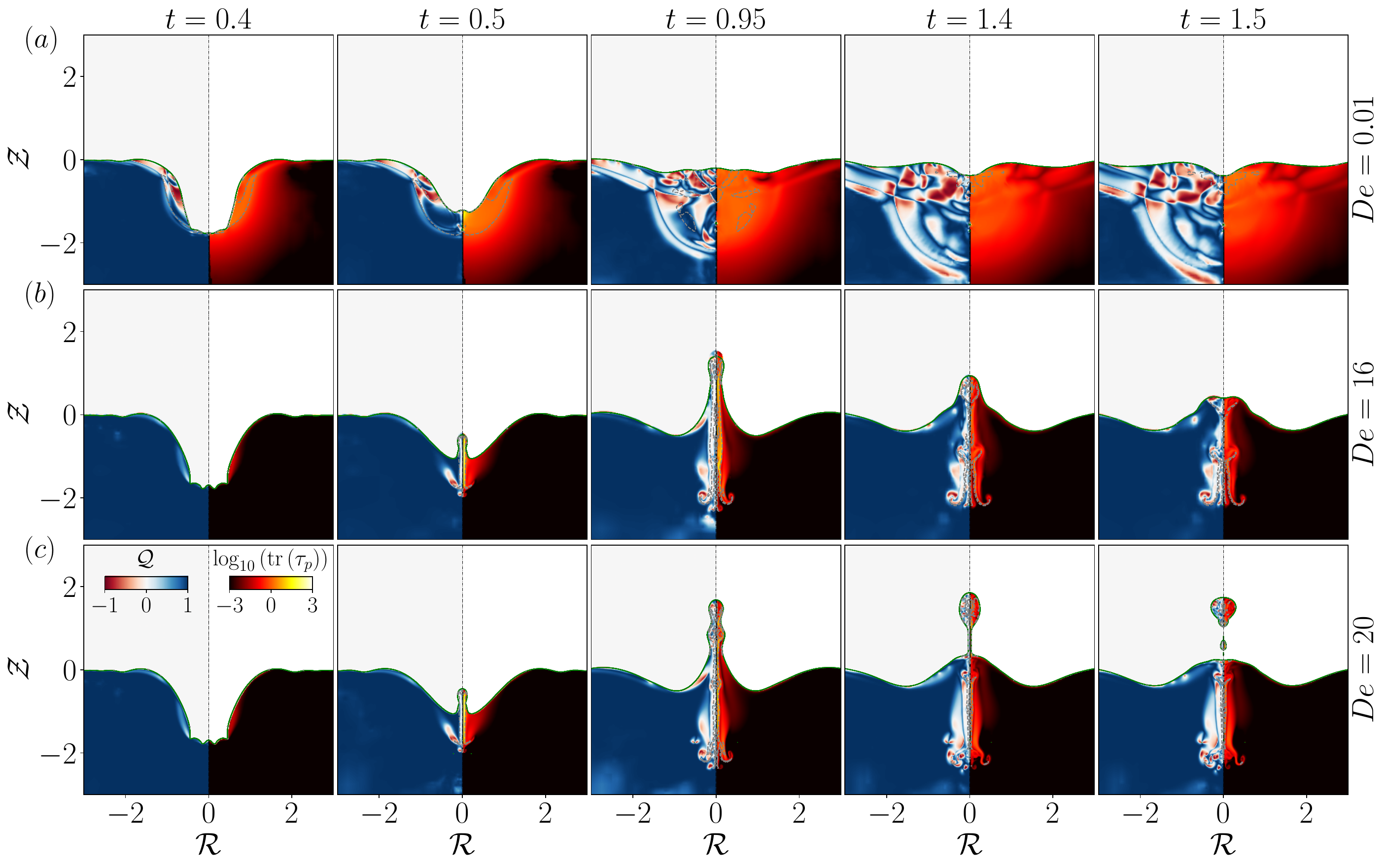}}
\caption[]{Variation of the dynamics of bubble bursting in an elasto-viscoplastic medium with respect to $\text{\De}$ at $\mathcal{J}=1$. $(a)$~$\text{\De}=0.01$, $(b)$~$\text{\De}=16$, $(c)$~$\text{\De}=20$. The left part of each panel shows the flow topology parameter $\mathcal{Q}$ and the right part of the panel shows the trace of elastic stress on a $\mathrm{log_{10}}$ scale. The yielded regions are marked by a grey line, which corresponds to~$\mathcal{J}\approxeq \|\boldsymbol{\tau_d}\|$.} 
\label{fig_index_plot_De_2}
\end{figure}

Bursting bubble in an EVP fluid exhibits a non-monotonic behaviour in the jet development process, as shown in figure~\ref{fig_index_plot_De_1}$b-d$. The figure illustrates three representative cases, identifying the effects of elastic stress relaxation on the bubble bursting process in an EVP fluid (supplementary movies are available at~\cite{ariGithub}) for a given plastocapillary number of~$\mathcal{J}=0.1$. The left and right panels of figure~\ref{fig_index_plot_De_1}$b-d$ correspondingly show the flow topology parameter~$\mathcal{Q}$ and the trace of elastic stresses on a $\mathrm{log_{10}}$ scale. Note that the early-time dynamics ($t \lessapprox 0.4$) appear similar in figure~\ref{fig_index_plot_De_1}$b-d$, but exhibits qualitatively different behaviour at later times. At low~$\text{\De}$, droplet formation is observed which is eventually suppressed by elastic stresses for intermediate values of~$\text{\De}$. For large relaxation time (\textit{i.e.} at higher $\text{\De}$), the elastic stresses persist and are concentrated close to the jet. 
\oo The jet is characterized by larger axial stress in the elongational flow region of the jet (characterized by high shear $\mathcal{Q}\approx 0$), and extra stress opposes the capillary stress, inhibiting the droplet formation from the jet\bb. This observation of droplet prevention due to the addition of polymers was also discussed by~\cite{rodriguez2023,dixit2024}. At even higher values of~$\text{\De}$~(\textit{i.e.} $\text{\De}=20$), we observe again a persistent Worthington jet that thins appreciatively over time (see figure~\ref{fig_index_plot_De_1}$c$).
\oo Due to low elasto-capillary number, the bulk medium remains unyielded and behaves as a Kelvin--Voigt viscoelastic solid with negligible~$\mathrm{tr}(\boldsymbol{\tau_p})$. In contrast, the axial region of the jet experiences significant extra stress, and as the jet continues to thin, it eventually pinches off when its thickness becomes smaller than the grid size\bb.


It is essential to note that the appearance of the yielded region ($K \neq 0$) is influenced not only by the plasticity of the fluid but also by the elastic stress ($\tau_p$), which affects the yield criterion (refer equation~\eqref{eqn:Aconform}). For~$\mathcal{J}=0.1$, at low $\text{\De}$, we observe a large yielded region exhibiting significant elastic deformation from base state (\textit{i.e.} $A \neq I$). Hence, the resulting elastic stresses relax more rapidly due to shorter relaxation times. This fast extra stress relaxation, especially at low~$\text{\De}$, causes the EVP medium to behave similarly to a Newtonian fluid, as evidenced by the similar busting bubble dynamics. However, it is worth noting that the resulting jet formation differs from the Newtonian case at~$\text{\Oh}=10^{-2}$ as the introduction of yield stress increases the apparent viscosity (see~\S\ref{subsec:vatsal_validation}) of the fluid and alters the flow. Note that the maximum magnitude of elastic deformation occurs mainly in the region of high shear around the time when capillary waves converge at the bottom of the bubble cavity , resulting in jet formation. With increasing $\text{\De}$ at fixed $\text{\Oh}_p$, the elastocapillary number decreases, indicating lower elastic energy in the EVP fluid~(refer~\S\ref{subsec:energy_budget}) and thereby, the yielded region appears in the proximity of the jet. The elastic stresses are lower in the majority of the bulk, which remains unyielded ($K = 0$) and exhibits a Kelvin--Voigt viscoelastic solid behaviour ($\boldsymbol{\overset{\nabla}{A}} = \bold{0}$). (Note that the higher elastic stresses can be found in the yielded region of the EVP fluid where $\mathrm{tr}(A)$ can be larger.) At high $\text{\De}$, a significant portion of the EVP fluid behaves as an elastic solid with very low elastic modulus, which allows the development of a high and thin jet, leading to capillary-type instability and pearl or drop formation. At $\text{\De} = 20$ (refer figure~\ref{fig_index_plot_De_1}$c$), we observe jet formation reminiscent of the \emph{bead-on-a-string} instability, which has been suggested to be the elastic counterpart of Rayleigh-Plateau instability, and is observed in both viscoelastic fluids and elastic solids~\citep{kibbelaar2020}.

The effects of increasing plasticity (via plasto-capillary number~$\mathcal{J}$) are shown in figure~\ref{fig_index_plot_J}, at a small Deborah number~($\text{\De}=0.04$). We observe that with the addition of plasticity, jet formation is monotonically suppressed. This is due to the increased apparent viscosity, as discussed in~$\S$\ref{subsec:vatsal_validation}. 

For a lower $\text{\De}$, the elastic stresses relax quickly, and the material behaves as a viscous fluid when the elastic stresses are negligible. Here, for the considered~$\text{\De}\sim \mathcal{O}(0.01)$, the elastocapillary number is higher, indicating that most of the EVP fluid region is yielded and the magnitude of elastic energy is high as observed from figure~\ref{fig_index_plot_J}$a-c$. Then, for a lower~$\mathcal{J}$, $K \neq 0$ over significant portion of the EVP fluid, indicating a larger yielded region around the axis of cavity and jet. With an increase in the plasticity of the elasto-viscoplastic fluid, the yielded region decreases for a given $\text{\De}$. However, as mentioned earlier, it is essential to note that the elastic stress also plays a role in influencing the yield criterion $K$.

Finally, we examine the combination of moderate to considerable plasticity~($\mathcal{J}=1$) and different relaxation time of elastic stresses (see figure~\ref{fig_index_plot_De_2}). The behaviour is qualitatively similar to that observed at~$\mathcal{J} = 0.1$. However, at low~$\text{\De}$~(figure~\ref{fig_index_plot_De_2}$a$), jet formation is wholly suppressed, indicating that the EVP fluid takes on the characteristics of a viscoplastic fluid. The deformation of the bubble cavity is significantly dampened by both the yielded region (with increased apparent viscosity) and the unyielded region (with high elastic modulus because of high $Ec$). For~$\text{\De}\ge 1$, the jet is pronounced due to lower elastic energy than capillary energy~(low elastocapillary numbers). Further, for~$\text{\De}=20$ at~$t=0.95$ the growth of elastic instability compensated by the surface tension forces in the jet hints at the appearance of~\emph{bead-in-a-string} characteristic. At medium~$\text{\De}$ however, the elastic stresses and plasticity acts together to broaden and suppress the jet formation. In the forthcoming subsections, we will scrutinize the effect of elastoviscoplastic rheology on each stage of the bubble bursting process.

\subsection{Capillary waves in the presence of yield stress and elastic stress} \label{subsec:capillary_wave}

\begin{figure}
\centering
\resizebox*{0.95\linewidth}{!}{\includegraphics{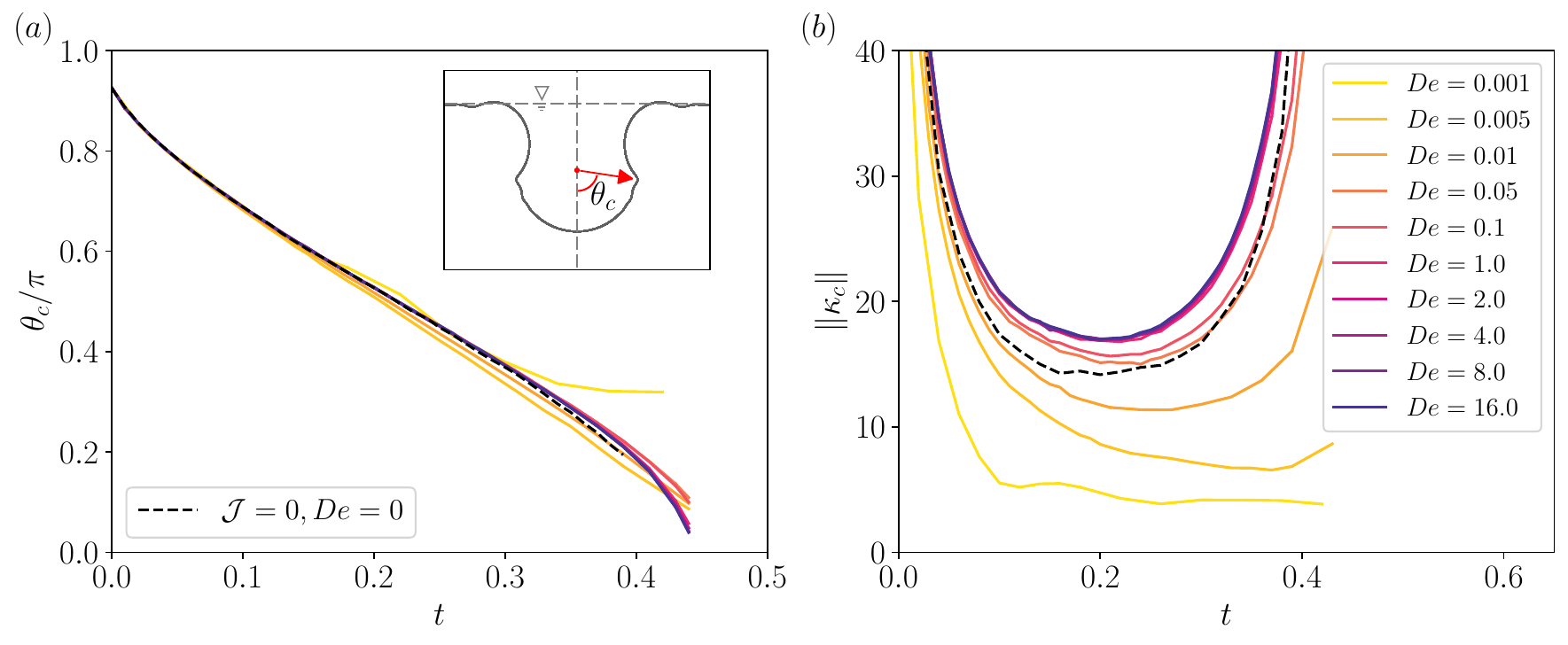}}
\caption[]{Effects of dimensionless relaxation time on the travelling capillary waves at~$\mathcal{J}=1$. $(a)$~Time variation of the location of strongest capillary wave $(\theta_c)$~for different $\text{\De}$. $(b)$~Corresponding time variation of the strength of strongest capillary wave $(\|\kappa_c\|)$.
} 
\label{fig_theta_kappa_De}
\end{figure}

Capillary waves are critical in the bubble-bursting process~\citep{gordillo2019,vatsal2021}. Initially, as the film breaks and the rim retracts, they create a series of capillary waves with varying strengths~\citep{gekle2009}. However, the high-frequency capillary waves experience substantial viscous damping. Consequently, the process of wave focusing and jet formation is controlled by the strongest wave, which remains unimpeded by viscous damping. Here, we track the strongest wave by monitoring the maximum curvature of the free-surface wave~\hlrev{$(\|\kappa_c\|$, see inset of figure~\ref{fig_theta_kappa_De}$a$ for the location~($\theta_c$) of the strongest capillary wave)}, as in~\cite{vatsal2021}, to understand the combined role of elastic stress relaxation and plasticity in jet formation.

\begin{figure}
\centering
\resizebox*{0.95\linewidth}{!}{\includegraphics{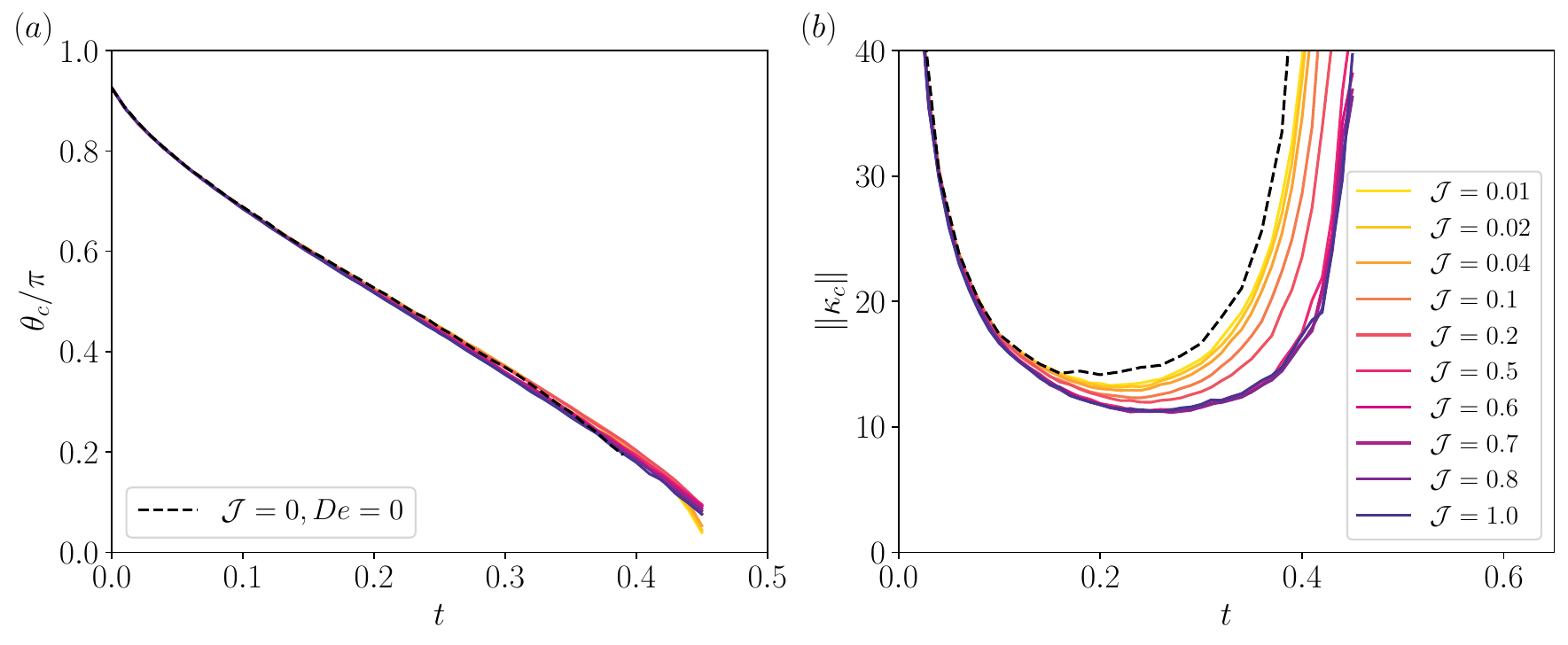}}
\caption[]{Effects of plasticity on the travelling capillary waves at~$\text{\De}=0.01$. $(a)$~Time variation of the location of strongest capillary wave $(\theta_c)$ for different $\mathcal{J}$. $(b)$~Corresponding time variation of the strength of strongest capillary wave $(\|\kappa_c\|)$.  
} 
\label{fig_theta_kappa_J}
\end{figure}

The location and amplitude of the strongest capillary wave as a function of the Deborah number are shown in figure~\ref{fig_theta_kappa_De} (for $\mathcal{J}=1$). In the case of Newtonian fluid, the capillary wave travels with a constant inertio-capillary velocity~$(\sim V_\gamma)$ as shown in figures~\ref{fig_theta_kappa_De}$a$ (dashed line). However, the viscous stress impedes these waves, resulting in the decrease of the strength of the wave~($\|\kappa_c\|)$, as observed in figure~\ref{fig_theta_kappa_De}$b$. For~$\theta_c/\pi \approx 0.5$, the cavity geometry changes, leading to flow focusing where an increase in strength of the capillary wave is observed.

From figure~\ref{fig_theta_kappa_De}, we observe that the location of the strongest wave is not significantly affected by elastic stresses in the EVP fluid. Consequently, the strongest elasto-capillary waves still travel with the inertio-capillary velocity~$V_\gamma$, due to the high shear yielded region close to the cavity interface, which behaves as a viscoelastic fluid. However, at low~$\text{\De}$, as the material approaches the viscoplastic limit, the location of the strongest wave tends to deviate from that of the Newtonian fluid, particularly in the flow-focusing part around~$t\approx 0.3$ owing to enahcned apparent viscosity. At low~$\text{\De}$, the strength of the capillary wave is also notably reduced, as evident in figure~\ref{fig_theta_kappa_De}$b$, due to both viscous and polymer dissipation~(see also~\S\ref{subsec:energy_budget}). The combined dissipation mitigates the jet formation. The increased polymeric dissipation is due to the faster relaxation of the elastic stresses close to the free surface where the fluid is yielded. Conversely, for high~$\text{\De}$, the elastocapillary number decreases, and the corresponding deformation of the cavity results in the presence of a stronger capillary wave than in the Newtonian case.


We now investigate the role of plasticity in the propagation of capillary waves at low~$\text{\De}$. In figure~\ref{fig_theta_kappa_J}$a$, we observe that the elasto-capillary waves travel with the same inertio-capillary velocity~$V_\gamma$ and show no deviation from the Newtonian case for the range of considered plasto-capillary numbers in this study. At low~$\text{\De}$, $\mathcal{J}$ primarily influences the yield-criterion ($K$). For the~$\mathcal{J}$ range in this study, the region close to the cavity interface is yielded and behaves as a viscoelastic fluid. Consequently, the propagation of capillary waves is unaffected by an increase in $\mathcal{J}$. However, an increase in $\mathcal{J}$ increases the apparent viscosity leading to attenuation of the strongest capillary wave and mitigation of the Worthington jet, following the mechanism proposed by \citet{gordillo2019} and \citet{vatsal2021} for Newtonian and purely Bingham fluids, respectively (figures~\ref{fig_theta_kappa_J}$b$ and~\ref{fig_index_plot_J}).

The EVP fluid studied here feature a notable distinction from the purely Bingham fluids. \citet{vatsal2021} noted that the damping of capillary waves leads to the retention of surface energy, culminating in non-flat final shapes. This phenomenon contrasts with the behavior of Elasto-ViscoPlastic (EVP) fluids. In EVP fluids, the high elastic modulus imparts solid-like characteristics, particularly in the context of elastic stress relaxation. During the cavity deformation process, these stresses accumulate and only dissipate gradually. This slow relaxation is responsible for the eventual deformation of the cavity over time depending on elastocapillary number, a stark difference from the Bingham fluid case where such deformation is impeded by the stored surface energy.

\subsection{Influence of elastic stress relaxation time and yield stress on jet formation}

\begin{figure}
\centering
\resizebox*{0.95\linewidth}{!}{\includegraphics{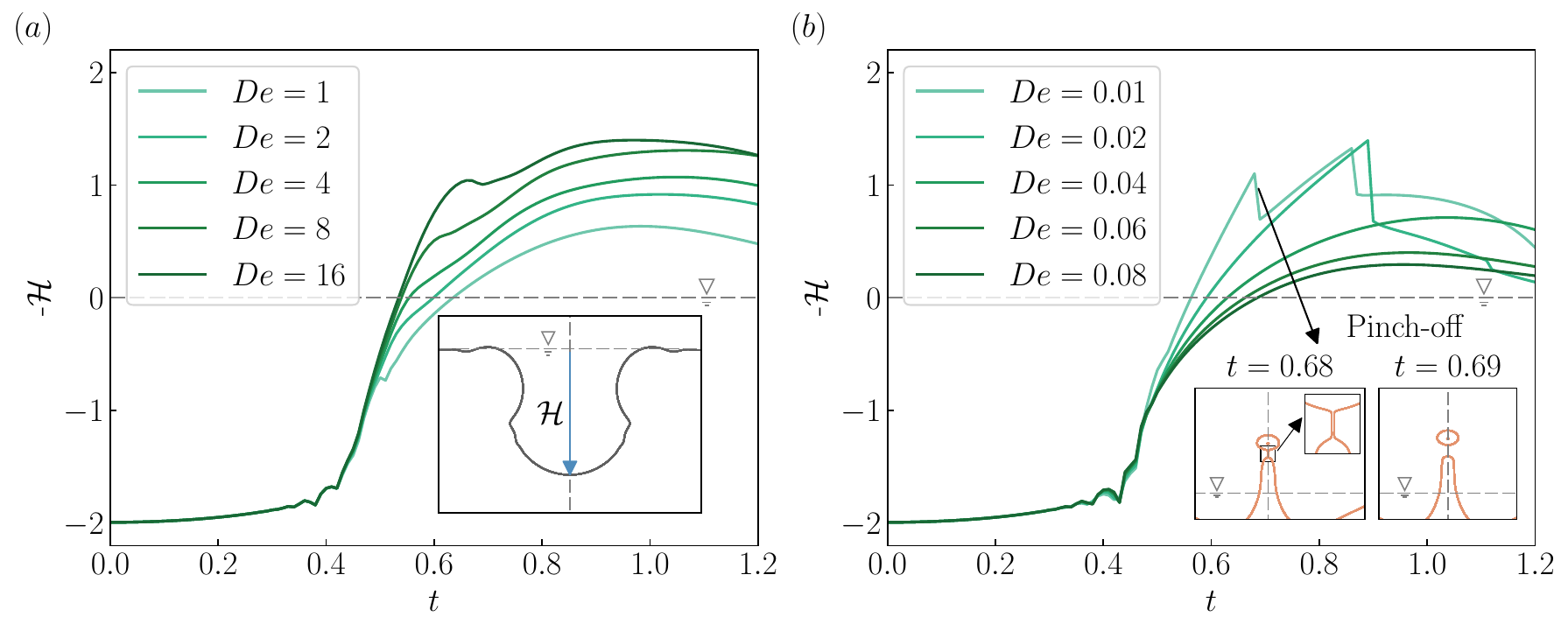}}
\caption[]{Effects of non-dimensional elastic stress relaxation time on the formation of jet as a result of the collapsing cavity. Variation
of the depth $\mathcal{H}$ of the cavity at its axis with time for (left) high Deborah numbers and (right) for low Deborah numbers at $\mathcal{J}=0.1$. (a--inset) shows the definition of $\mathcal{H}$ and (b--inset) depicts the pinch-off process. 
} 
\label{fig_height_De}
\end{figure}

\begin{figure}
\centering
\resizebox*{0.95\linewidth}{!}{\includegraphics{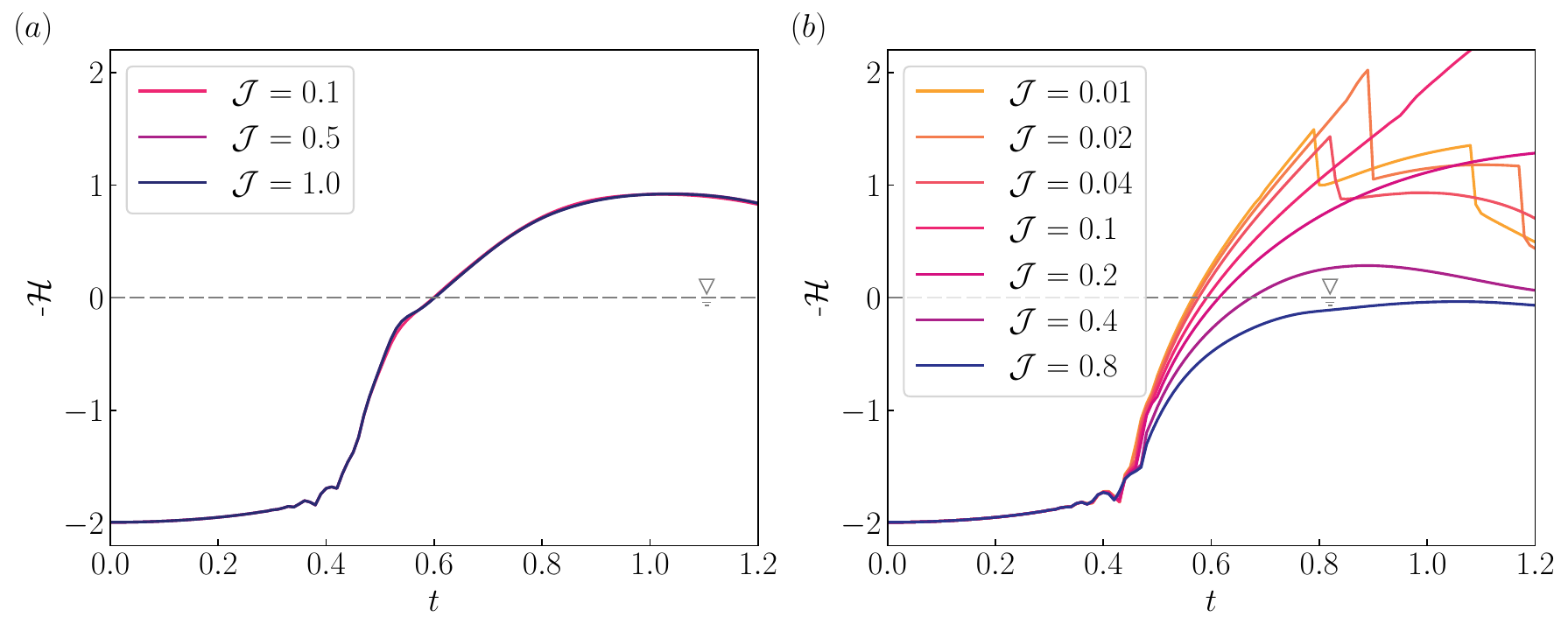}}
\caption[]{Effects of plasticity on the formation of jet as a result of the collapsing cavity. Variation
of the depth $\mathcal{H}$ of the cavity at its axis with time at (left)~$\text{\De}=2$ and (right)~$\text{\De}=0.02$.
} 
\label{fig_height_J}
\end{figure}

One of the key characteristics of the bubble-bursting process is the formation of the Worthington jet~($-\mathcal{H}>0$ at $\mathcal{R} = 0$) following the collapse of the bubble cavity. Figure~\ref{fig_height_De} illustrates the temporal evolution of the jet for different~$\text{\De}$. From figure~\ref{fig_height_De}a, an increase in $\text{\De}$ of the EVP fluid pronounces the jet growth and the maximum height of the jet. For~$\text{\De} \sim \mathcal{O}(1)$, the elastic stresses in the jet do not relax and compensate the capillary stresses resulting in the prevention of droplet break-up from the jet.

However, from figure~\ref{fig_height_De}$b$, for~$\text{\De} \le 0.04$, we observe the droplet formation as the elastic stresses (also relaxing faster) are overcome by the capillary forces. However, it is intuitive to note that as Deborah number increases for~$\text{\De}\sim \mathcal{O}(0.01)$, the maximum jet height decreases. The same behaviour is also observed at~$\mathcal{J}=1$ (not shown here due to brevity) where the droplet pinch-off is not observed. The decrease in jet development at low but increasing~$\text{\De}$ (as shown in figure~\ref{fig_height_De}$b$) is due to relatively slower relaxation of elastic stresses and thereby increased storage of elastic energy at the time when capillary waves collapse at the bottom of the cavity, i.e. during jet initiation~(refer Appendix~\ref{appendix:jet_growth}). Apart from this, combined effects of increasing viscous dissipation and decreasing polymeric dissipation also contributes to the behaviour of jet development. However, as~$\text{\De}$ increases (for~$\text{\De}\sim \mathcal{O}(1)$ as highlighted in figure~\ref{fig_height_De}$a$), the elastocapillary number decreases indicating lower elastic energy, which pronounces the jet height~(refer Appendix~\ref{appendix:jet_growth}).

Figure~\ref{fig_height_J} reveals that for~$\text{\De}\sim \mathcal{O}(1)$~(\textit{i.e.} in the visco-elasto-capillary regime), plasticity does not appear to have a significant impact on jet development. However, as the relaxation time of the elastic stress decreases in the EVP fluid, the influence of plasticity becomes more pronounced, resulting in the reduction of jet growth.

\begin{figure}
\centering
\resizebox*{0.5\linewidth}{!}{\includegraphics{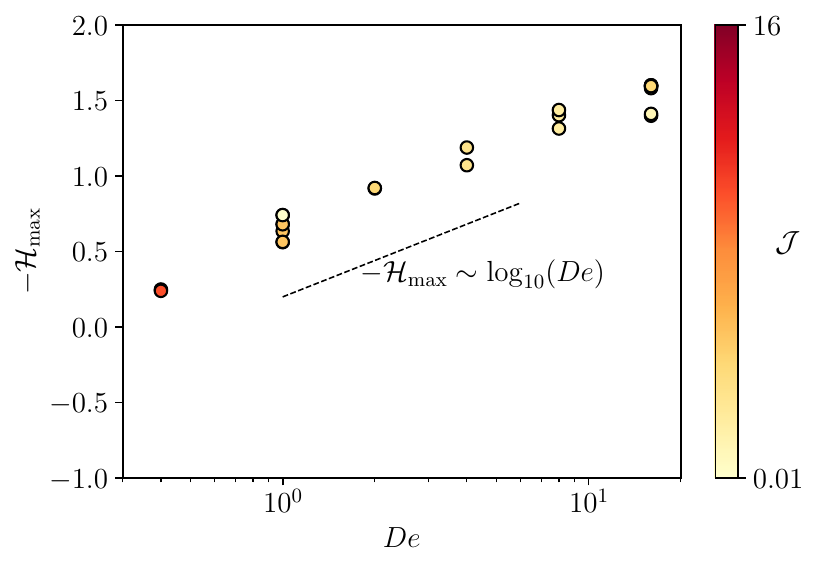}}
\caption[]{Effects of elastic stress relaxation on the maximum height of Worthington jet formation. The darker markers correspond to higher~$\mathcal{J}$ and vice versa.
} 
\label{fig_max_jet_height}
\end{figure}

\begin{figure}
\centering
\resizebox*{0.95\linewidth}{!}{\includegraphics{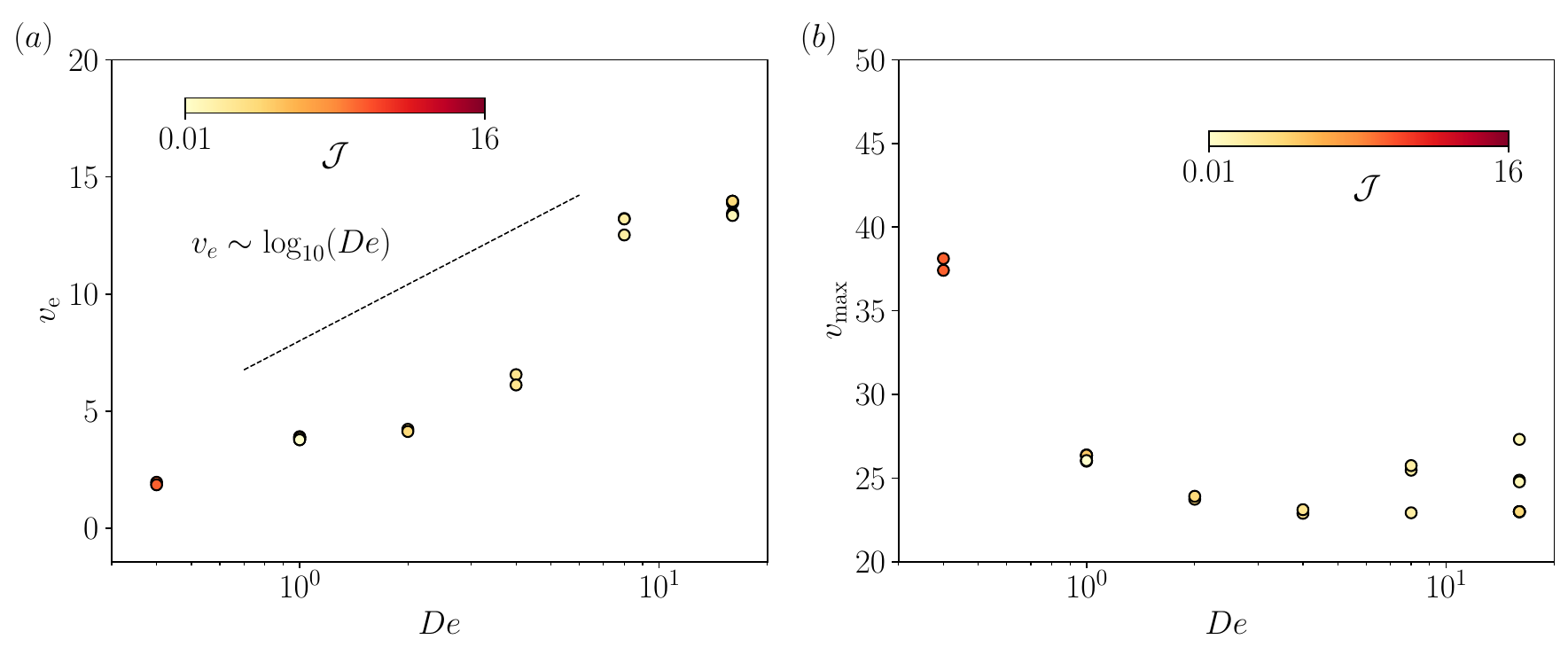}}
\caption[]{Variation of the Worthington jet velocity with dimensionless elastic stress relaxation time. (Left) Velocity of the jet at the instant when the interface crosses the equilibrium surface $(v_e)$ is plotted against $\text{\De}$. (Right) The effects of elastic stress relaxation on the maximum velocity $(v_\mathrm{max})$ of the jet. The darker markers correspond to higher~$\mathcal{J}$ and vice versa.
} 
\label{fig_jet_velocity}
\end{figure}

Figure~\ref{fig_max_jet_height} illustrates the variation in the maximum jet height~$(-\mathcal{H}_{\mathrm{max}})$ with respect to dimensionless relaxation time of the elastic stresses in the EVP fluid. In the visco-elasto-capillary regime, the trend with which the maximum height of the jet varies is logarithmic with respect to~$\text{\De}$.

In Figure~\ref{fig_jet_velocity}, we present the variation of jet velocity in relation to~$\text{\De}$. In this context, the jet velocity is defined as the vertical velocity of the interface at the axis. We use~$v_e$ to represent the velocity of the jet as it crosses the free-flat equilibrium surface, specifically~$v_e = v(\mathcal{Z}=\mathcal{R}=0)$. Note that~\cite{ghabache2014} have utilized a similar quantification of jet velocity in the experiments of bubble bursting in a Newtonian medium. Additionally, we also sample~$v_{\mathrm{max}}$, which corresponds to the maximum vertical velocity of the interface along the axis~\citep{gordillo2019,vatsal2021,Gordillo2023}. Typically, this maximum velocity occurs when capillary waves merge at the bottom of the bubble cavity. We hypothesize that the maximum interface velocity~$v_{\mathrm{max}}$ at the axis represents the initiation velocity of the jet. This initiation velocity results from the conversion of capillary surface energy (due to merging capillary waves) into the kinetic energy of the jet.

From figure~\ref{fig_jet_velocity}$a$, we observe the jet velocity at the equilibrium surface almost exhibits a logarithmic increase with~$\text{\De}$. Furthermore, from figure~\ref{fig_jet_velocity}$b$, we notice that the initiation velocity of the jet remains relatively constant with~$\text{\De}$ within the visco-elasto-capillary regime. Consequently, the maximum kinetic energy, which arises from the energy balance at the moment of capillary wave merging at the bottom of the cavity, is minimally affected due to a decrease in the elastocapillary number in the visco-elasto-capillary regime.

\begin{figure}
\centering
\resizebox*{0.5\linewidth}{!}{\includegraphics{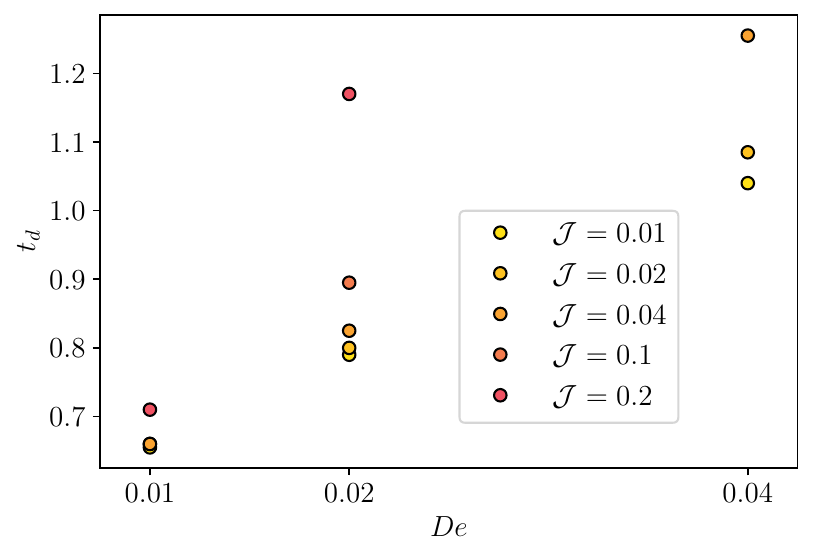}}
\caption[]{Variation of the droplet-formation time with respect to dimensionless relaxation time of elastic stress.  
} 
\label{fig_drop_time}
\end{figure}

Jet formation can lead to droplet formation at low and high Deborah numbers. In this study, we qualitatively discuss the impact of elastic stress relaxation time on the jet's droplet-formation time~($t_d$) in the jet using figure~\ref{fig_drop_time}. However, a quantitative analysis of droplet formation time is not accurate in this study due to the numerical artefact, which is detailed in~Appendix~\ref{appendix:grid_independence}. In the near-Newtonian limit characterized by a lower relaxation time, we observe that the droplet formation time is delayed logarithmically with increasing~$\text{\De}$. This delayed droplet–formation
is attributed to a finite non-zero relaxation time of elastic stresses, which increases with $\text{\De}$ and thereby delays the time at which the capillary stresses overtake the elastic stresses in the jet.

\subsection{Effects of solvent-polymer viscosity ratio on Worthington jet} \label{subsec:beta_variation}

\begin{figure}
\centering
\resizebox*{0.95\linewidth}{!}{\includegraphics{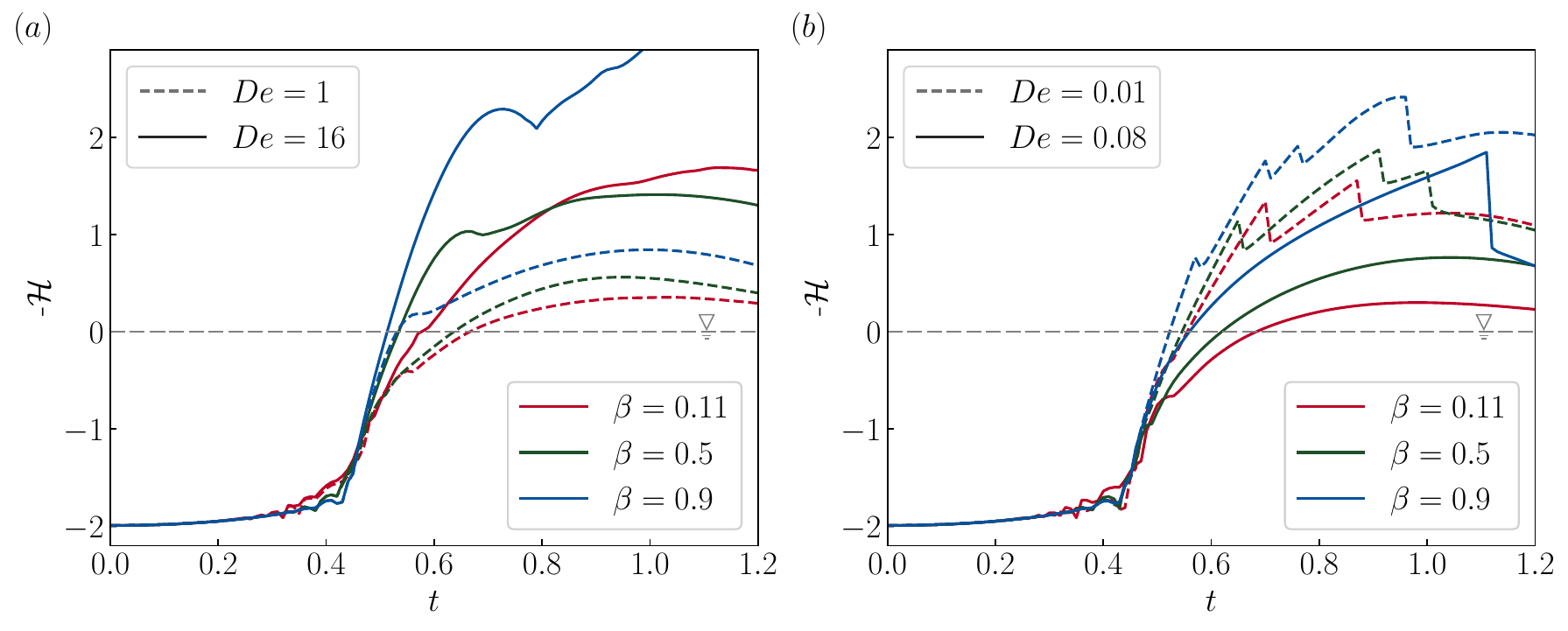}}
\caption[]{Effects of solvent-polymer viscosity ratio on jet evolution. Variation
of the depth $\mathcal{H}$ of the cavity at its axis with respect to time for ($a$) $\mathcal{J}=1$ and ($b$) $\mathcal{J}=0.01$.
} 
\label{fig_height_beta}
\end{figure}

\begin{figure}
\centering
\resizebox*{0.85\linewidth}{!}{\includegraphics{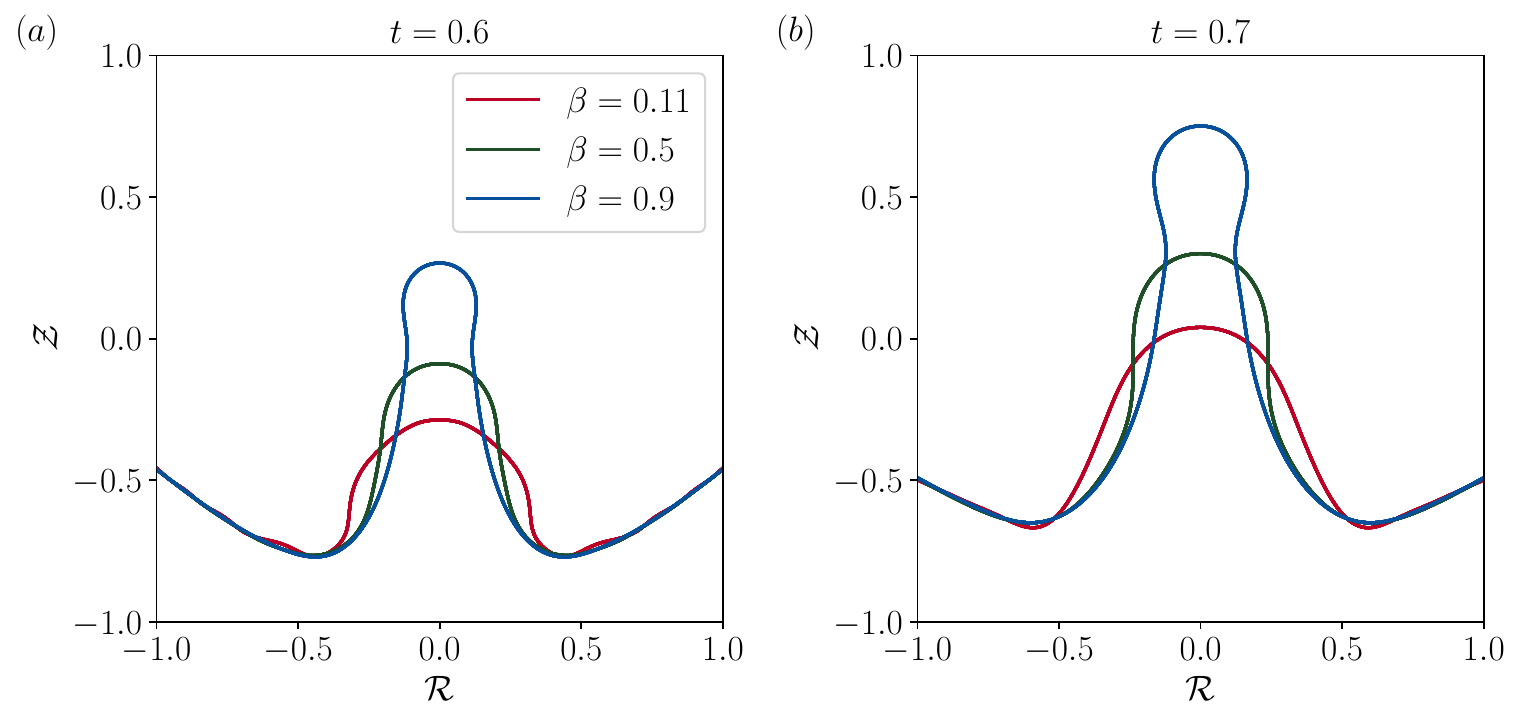}}
\caption[]{Variation of jet thickness with respect to solvent-polymer viscosity ratio quantified by~$\beta\, (=\text{\Oh}_s/(\text{\Oh}_s+\text{\Oh}_p))$ at (a)~$t=0.6$, (b)~$t=0.7$ for $\mathcal{J}=0.01$ and $\text{\De}=0.08$.  
} 
\label{fig_jet_thickness}
\end{figure}

In figure~\ref{fig_height_beta} we show the effects of varying solvent-polymer viscosity ratio in terms of~$\beta\,(=\text{\Oh}_s/(\text{\Oh}_s+\text{\Oh}_p))$ on the jet development at different~$\text{\De}$ in the visco-elasto-capillary regime~(figure~\ref{fig_height_beta}$a$) and at lower~$\text{\De}$~(figure~\ref{fig_height_beta}$b$). Overall, we observe that the jet development is pronounced in the visco-elasto-capillary regime at~$\text{\Oh}_s=0.009,\,\text{\Oh}_p = 0.001$~($\beta = 0.9$) in comparison to~$\text{\Oh}_s=0.0011,\,\text{\Oh}_p = 0.0088$~($\beta = 0.11$). Previously, we observed that relatively decreasing polymeric dissipation in the jet accentuates the growth of the jet in the visco-elasto-capillary regime~(refer also~Appendix~\ref{appendix:jet_growth}). Additionally, increasing~$\text{\Oh}_p$~($\beta \rightarrow 0$) increases the elastocapillary number, which indicates a higher elastic stress in the jet, which on relaxation dissipates energy and thereby reduces the kinetic energy of the jet.

We also observe a similar trend of pronounced jet development in the near-Newtonian limit, characterized by low~$\text{\De}$ and~$\mathcal{J}$, as shown in figure~\ref{fig_jet_thickness}. It is important to note that a similar trend of decreasing jet growth with increasing~$\text{\Oh}_p$ was also observed for lower~$\text{\De}$ at higher~$\mathcal{J}$, although it is not presented here. In summary, we find that increasing~$\text{\Oh}_p$ relative to~$\text{\Oh}$ (where~$\text{\Oh} = \text{\Oh}_s + \text{\Oh}_p$) has a similar effect to an increased apparent extensional viscosity in jet development. Additionally, when~$\beta$ decreases, the EVP fluid exhibits significant extensional viscosity, resulting in a thicker jet as evident in figure~\ref{fig_jet_thickness}.

\hlrev{With the above results, we probe into the modification of the regime map as plotted in figure~\ref{fig_regime_map} with respect to~$\beta$. In figure~\ref{fig_height_beta}$b$, note that we are in the droplet formation regime when~$\text{\De} = 0.01$. At this specific parameter point~($\text{\De}$, $\mathcal{J}$), different~$(\beta)$ exhibit droplet formation behavior. However, when we increase ~$\text{\De}$ from $0.01$ to $0.08$~(crossing the green boundary along~$\text{\De}$ from the droplet formation regime to the no-pinch-off regime in figure~\ref{fig_regime_map}), we observe that droplet formation is hindered for~$\beta = 0.5$. It is important to note that in figure~\ref{fig_height_beta}$b$, droplet formation is indicated by the discontinuity in~$\mathcal{H}$~(refer inset in figure~\ref{fig_height_De}$b$). For~$\beta=0.9$, however, droplet formation persists at a higher~$\text{\De}$, suggesting that the (green) boundary separating the droplet formation and no-pinch-off regimes shifts upwards to accommodate a higher~$\text{\De}$ in the case of higher~$\text{\Oh}_s$. Conversely, as $\text{\Oh}_p$ increases, the boundary would move closer to a lower~$\text{\De}$.

Considering figure~\ref{fig_height_beta}$a$, we can examine the transition boundary between the no-jet and no-pinch-off regimes (orange boundary along~$\text{\De}$ in figure~\ref{fig_regime_map}). For~$\text{\De}=1$, we observe that the maximum jet height decreases with increasing~$\beta$. This implies that the transition boundary (orange boundary along~$\text{\De}$) shifts downward as~$\beta$ increases and upwards as~$\beta$ decreases.

Finally, we turn our attention to the transition between the droplet formation regime and the no-jet regime for~$\text{\De} \sim \mathcal{O}(10^{-2})$ (which corresponds to the green boundary along~$\mathcal{J}$ in figure~\ref{fig_regime_map}). From equation~\eqref{eqn:apparent_viscosity_J}, the apparent viscosity increases with decreasing~$\beta$, resulting in the transition from droplet formation to no-pinch-off occurring at lower~$\mathcal{J}$ for decreasing~$\beta$. Consequently, this shifts the (green) boundary (along~$\mathcal{J}$) to lower~$\mathcal{J}$, as~$\text{\Oh}_p$ increases.}

\subsection{Energy analysis} \label{subsec:energy_budget}

\begin{figure}
\centering
\resizebox*{0.48\linewidth}{!}{\includegraphics{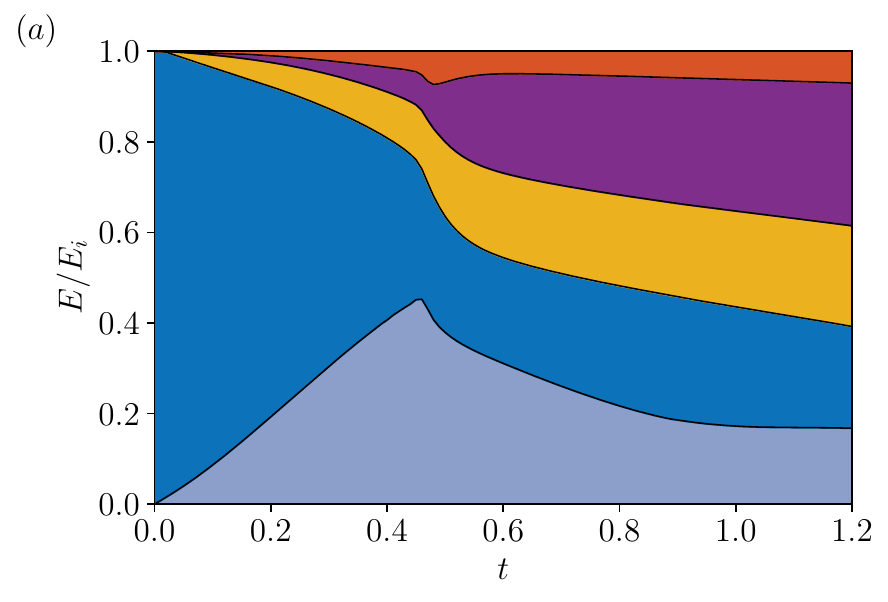}}
\resizebox*{0.48\linewidth}{!}{\includegraphics{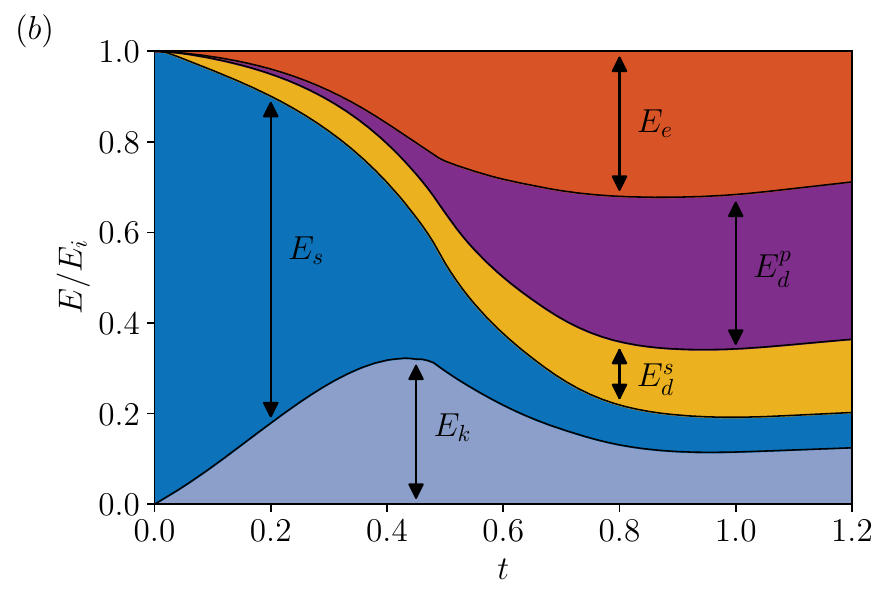}}
\resizebox*{0.48\linewidth}{!}{\includegraphics{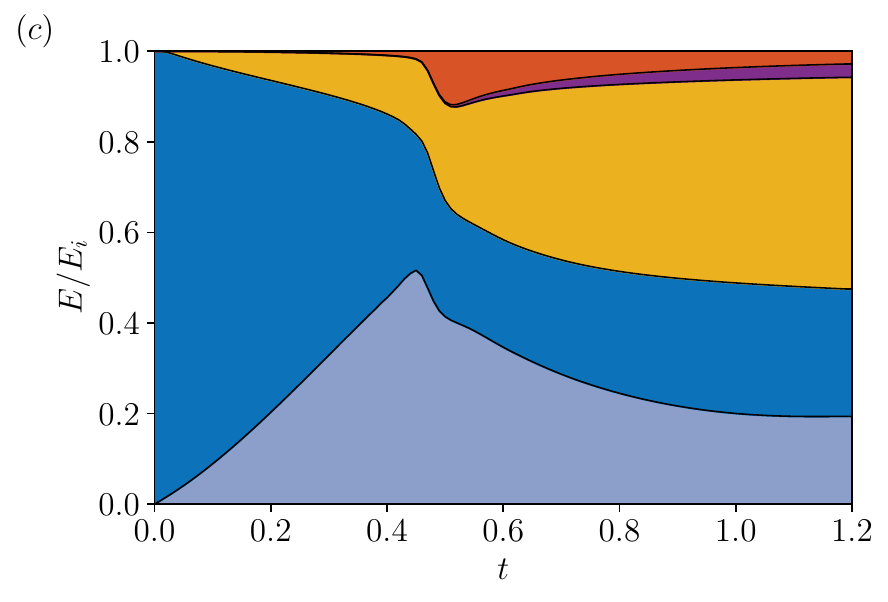}}
\resizebox*{0.48\linewidth}{!}{\includegraphics{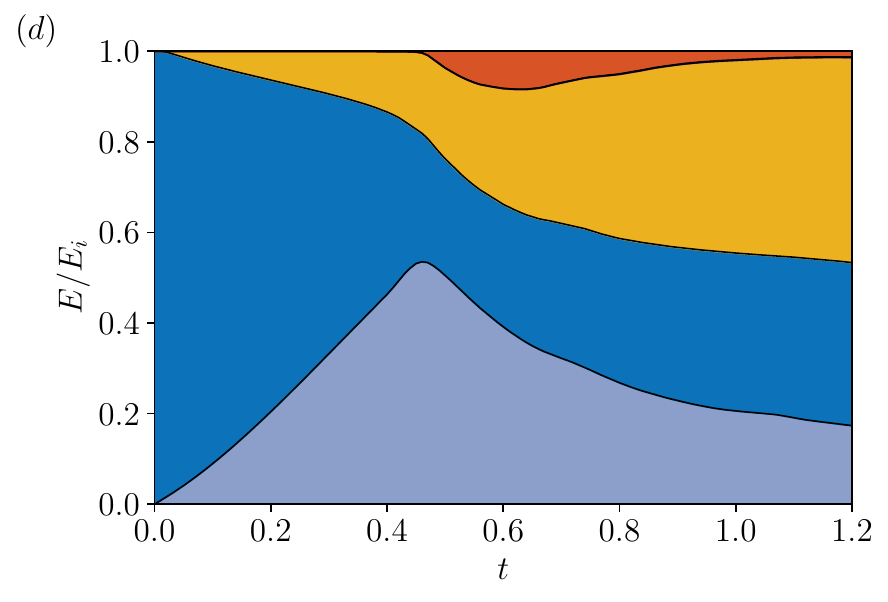}}

\caption[]{Energy budget for the bubble bursting in an elasto-viscoplastic medium. The plots indicate the variation of different modes of energy transfer with respect to time for $(a)$~$\mathcal{J}=0.1,\;\text{\De}=0.02$, $(b)$~$\mathcal{J}=1,\;\text{\De}=0.01$, $(c)$~$\mathcal{J}=0.1,\;\text{\De}=1$, $(d)$~$\mathcal{J}=1,\;\text{\De}=20$.  
} 
\label{fig_energy_budget}
\end{figure}

The total energy in the computational domain at any time $t$ is composed of kinetic energy~$(E_k(t))$, surface energy~($E_s(t)$, assuming flat free-surface to have zero surface energy), elastic energy~$(E_e)$ and, as well as energy dissipation due to both solvent viscosity~$(E_d^s(t))$ and polymer relaxation~$(E_d^p(t))$. The miscellaneous energy due to the motion of air, jet break-up, etc. is represented by~$E_m(t)$. 
The time-evolution of the energy in the system is expressed as:
\begin{align}
    E = E_k(t) + E_s(t) + E_e(t) + E_d^s(t) + E_d^{p}(t) + E_m(t)\,.
\end{align}

The system's initial energy at $t = 0$ consists entirely of surface energy, $E_i = E_s(0)$. We extend the energy analysis for Oldroyd-B viscoelastic fluids by \cite{snoeijer2020} to derive the energy budget for the elasto-viscoplastic fluid governed by \cite{saramito2007}'s model. Appendix~\ref{appendix:energy_budget} provides a detailed derivation of this budget.

Figure~\ref{fig_energy_budget} illustrates the energy budgets for four representative cases, each normalized by the initial energy $E_i$. These cases correspond to different regimes in the plasto-capillary number $\mathcal{J}$ and Deborah number $\text{\De}$ parameter space, as discussed in \S\ref{subsec:regime_map}. The figure shows the temporal evolution of various energy transfer modes.
As the EVP fluid deforms, surface energy converts into kinetic energy. This deformation induces viscous dissipation due to the solvent viscosity component. Additionally, relaxing elastic stresses contribute to energy dissipation ($E_d^p(t)$). However, the EVP fluid can store some energy as elastic energy ($E_e(t)$), potentially recoverable later in the bubble-bursting process.

\subsubsection{Droplet formation regime, Droplet--A (Low elastic stress relaxation time and yield-stress limit, $\mathcal{J} \rightarrow 0,\,\text{\De}\rightarrow 0$)}


Figure~\ref{fig_energy_budget}$a$ depicts the temporal energy budget for low~$\text{\De}$ and $\mathcal{J}$ values. The collapse of the high-energy bubble cavity initiates a cascade of energy conversions characterizing the early stages of bubble bursting. As the cavity collapses and capillary waves propagate ($t \lessapprox 0.5$), the initial surface energy $E_s(0)$ is released, increasing the fluid's kinetic energy $E_k(t)$. Significant velocity gradients lead to viscous dissipation $E_d^s(t)$ due to solvent viscous stresses, while the short relaxation time of elastic stresses causes a progressive increase in $E_d^p(t)$. Initially, solvent viscous stresses dominate, resulting in higher viscous dissipation compared to polymeric dissipation and elastic energy, despite $\text{\Oh}_p=\text{\Oh}_s=0.5\text{\Oh}$.


$E_k(t)$ reaches its maximum when capillary waves focus at the cavity bottom, creating a high-velocity region. As these waves merge to form a jet, elastic energy $E_e(t)$ is briefly stored and rapidly released as $E_d^p(t)$ due to small $\text{\De}$. The EVP fluid consistently stores more elastic energy as the jet develops, attributed to large axial elastic stresses in the jet and extra stress in both yielded and unyielded regions (see figure~\ref{fig_index_plot_De_1}$b$). The jet-development process is characterized by substantial energy dissipation. Cumulative solvent viscous and polymer dissipation critically regulate jet growth and maximum height. The elastic energy at jet initiation is also crucial, as discussed in Appendix~\ref{appendix:jet_growth}.

To approach the final stoppage time of bubble bursting dynamics in an EVP fluid, the rate of change of different energy transfer modes must be negligible. However, at $t=1.2$, the surface energy remains non-zero and decreasing. Capturing a steady final state would require significantly longer simulation times, likely on the order of $t \sim \mathcal{O}(10)$, which is beyond the scope of this study. For low $\text{\De}$, viscoelastic solid-like deformation in the unyielded region will lead to elastic stress enervation, and the cavity is expected to approach a flat free interface.



\subsubsection{No-jet regime (Viscoplastic limit, $\mathcal{J} \gg 0,\,\text{\De}\rightarrow 0$)}

At extremely large $\mathcal{J}$ values, most of the domain remains unyielded. In this regime, \citet{vatsal2021} observed finite $E_s(t \to \infty)$ as the cavity shape remained invariant, accompanied by a decrease in kinetic energy and total viscous dissipation. In contrast, our study reveals that at low $\text{\De}$, the visco-elastic deformation of the unyielded EVP fluid maintains cavity motion (see figure~\ref{fig_index_plot_De_2}), resulting in a gradual decrease of surface energy over time. Figure~\ref{fig_energy_budget}$b$ exemplifies this behavior with a consistent interplay between and dominance of $E_e(t)$ and $E_d^p(t)$.
Figure~\ref{fig_energy_budget}$b$ also shows that at higher $\mathcal{J}$ and lower $\text{\De}$, the peak kinetic energy during capillary wave convergence reaches only $\sim30\%$ of the initial total energy, compared to $>40\%$ in other regimes. Further reduction in $\text{\De}$ (at sufficiently high $\mathcal{J}$ and fixed $\text{\Oh}$) would lead to even lower kinetic energy and the formation of a non-flat equilibrium shape characteristic of bubble bursting in viscoplastic fluids. This occurs due to a large elasto-capillary number $Ec = \text{\Oh}_p/\text{\De}$, resulting in a balance between elastic and surface stresses.

\subsubsection{No pinch-off regime (Visco-elasto-capillary limit, $\text{\De} \sim \mathcal{O}(1)$)}



In the visco-elasto-capillary regime, elastic stresses play a minimal role ($E_e(t) \approx E_d^p(t) \approx 0$, figure~\ref{fig_energy_budget}$c$) during the initial bubble bursting in EVP fluid exemplified with $\mathcal{J}=0.1$. Nevertheless, the strength of the strongest capillary wave increases with Deborah number, as detailed in~\S\ref{subsec:capillary_wave}. At the moment of capillary wave convergence, significant elastic energy storage and release occurs in the EVP fluid, facilitating jet formation. In this regime, elastic energy storage primarily originates from the Kelvin--Voigt viscoelastic solid region, sustaining elongational stress in the jet. As elastic stress in the jet's axial region intensifies and subsequently relaxes, figure~\ref{fig_energy_budget}$c$ reveals a consistent increase in polymeric dissipation beyond $t \approx \text{\De}$.

\subsubsection{Droplet formation regime, Droplet--B (Newtonian-like limit, $\text{\De}\gg 1$)}
\vs
The energy transfer modes in this limit (figure~\ref{fig_energy_budget}$d$) closely resemble those observed in the visco-elasto-capillary limit (figure~\ref{fig_energy_budget}$c$), with negligible elastic energy storage and polymeric dissipation ($E_e(t) \approx E_d^p(t) \approx 0$) during the initial bubble bursting in EVP fluid. 
Additionally, this regime 
exhibits minimal elastic energy and polymeric dissipation during jet formation as well. These characteristics align with the Newtonian-like limit described by \citet{oishi2023}. Notably, for sufficiently large Deborah numbers ($\text{\De}$), regardless of the plastocapillary number ($\mathcal{J}$), elastic stresses become insignificant due to vanishing elastocapillary number ($Ec$) at finite solvent and polymeric Ohnesorge numbers ($\text{\Oh}_s, \text{\Oh}_p$) (also see \S~\ref{sec:NewtLike}). 


In summary, this section examines the energy budget during bubble bursting in an elasto-viscoplastic (EVP) fluid across various regimes defined by the plasto-capillary number $\mathcal{J}$ and Deborah number $\text{\De}$. The analysis reveals a complex interplay between surface energy, kinetic energy, elastic energy storage, and dissipation mechanisms. In the droplet formation regime (low $\mathcal{J}$ and $\text{\De}$), initial surface energy rapidly converts to kinetic energy and dissipation, with solvent viscous stresses dominating early stages despite equal solvent and polymeric Ohnesorge numbers.
As $\mathcal{J}$ increases and $\text{\De}$ decreases, the no-jet regime emerges, characterized by lower peak kinetic energy and persistent cavity motion due to visco-elastic deformation. The visco-elasto-capillary limit ($\text{\De} \sim \mathcal{O}(1)$) exhibits minimal elastic effects initially, but significant elastic energy storage during jet formation. In the Newtonian-like limit ($\text{\De} \gg 1$), elastic effects become negligible throughout the process.
Across all regimes, the dynamics of an EVP fluid exemplify a strong $\text{\De}$-dependent relationship between viscous and polymeric dissipation. Viscous dissipation increases with $\text{\De}$, while polymeric dissipation becomes more significant at lower $\text{\De}$ values, proportional to $\text{\De}^{-2}$ (for fixed $\mathrm{tr}\left(\boldsymbol{A} - \boldsymbol{I}\right)$ and $\text{\Oh}_p$). The polymeric dissipation, given by
\begin{align}
E_d^p = \int_t\left(\int_\Omega \frac{K\text{\Oh}_p}{\text{\De}^2}\left(\mathrm{tr}\left(\boldsymbol{A} - \boldsymbol{I}\right)\right)d\Omega\right)dt,
\end{align}
is confined to the yielded region of the fluid ($K \ne 0$).
In contrast, elastic energy storage occurs in both yielded and unyielded regions. Solvent viscous dissipation $E_d^s$ also manifests in both regions; in the unyielded region, it arises from the Kelvin--Voigt solid behavior where $\|\boldsymbol{\mathcal{D}}\| \ne 0$, resulting in dissipation.
\bb

\section{Conclusions \& outlook}
\label{sec:conclusion}

In this study, we investigate the bubble-bursting phenomenon using direct numerical simulations in an elasto-viscoplastic (EVP) medium to understand the interplay of elasticity~(more particularly the elastic stress relaxation time characterized by $\text{\De}$) and plasticity~(identified by $\mathcal{J}$) on the bursting mechanism. We observed distinct behaviors in the development of the jet depending on the control parameters~$\text{\De}$ and~$\mathcal{J}$ of an EVP fluid, resulting in four main regimes:

\emph{(i) Droplet formation regime, Droplet--A (Low elastic stress relaxation time and yield-stress limit, $\mathcal{J} \rightarrow 0,\,\text{\De}\rightarrow 0$)}: 
This regime is characterized by a Newtonian-like jet growth and droplet formation. In this regime, due to the faster relaxation of elastic stresses, the resulting capillary stresses lead to the formation of a jet despite a higher elastocapillary number. Further, the fast decay of elastic stresses cannot suppress the droplet formation due to the Rayleigh-Plateau instability. In this Newtonian-like limit, increasing~$\text{\De}$ reduces the maximum jet height due to slower relaxation of elastic stresses and thereby an increase in the elastic energy at the time of jet initiation (see Appendix~\ref{appendix:jet_growth}).

\emph{(ii) No-jet regime (Viscoplastic limit, $\mathcal{J} \gg 0,\,\text{\De}\rightarrow 0$)}: 
In this regime, the elasto-viscoplastic (EVP) fluid behaves similarly to a viscoplastic fluid. This regime is characterized by the absence of jet formation due to the increased apparent viscosity, which dampens capillary wave motion and suppresses jet initiation. Although the fluid remains mostly unyielded, exhibiting Kelvin--Voigt viscoelastic solid behaviour, it can slowly deform over time depending on the elastocapillary number. This behaviour contrasts with purely viscoplastic fluids, where non-flat final equilibrium shapes are retained due to the stored surface energy because of rigid behaviour in the unyielded region.

\emph{(iii) No pinch-off regime (Visco-elasto-capillary limit, $\text{\De} \sim \mathcal{O}(1)$)}: 
In this regime, jet formation occurs without subsequent droplet generation. This regime is marked by a balance between elastic stresses and capillary forces, influenced by the slower relaxation time of elastic stresses. As a result, the jet grows consistently but does not break into droplets. Notably, plasticity has minimal impact on bubble bursting dynamics due to the lower elastocapillary number. The elastic energy and polymer dissipation decrease with increasing $\text{\De}$, leading to higher maximum jet heights, indicating that slower elastic stress relaxation contributes to more pronounced jet formation.

\emph{(iv) Droplet formation regime, Droplet--B (Newtonian-like limit, $\text{\De}\gg 1$)}: 
In this regime, the behaviour of the EVP fluid resembles a Newtonian-like fluid, with the jet eventually thinning and pinching off from its base. In this regime, most of the EVP fluid exhibits unyielded Kelvin--Voigt viscoelastic solid behaviour, 
\oo due to low elasto-capillary number\bb. 
\oo The low elastic stresses are unable to prevent the formation of the Worthington jet. The resulting jet experiences significant axial stress, which causes it to yield, leading to progressive thinning and eventual breakup at its base\bb. Notably, the absence of substantial EVP stress relaxation in this regime makes polymeric dissipation a negligible factor. The influence of solvent-polymer viscosity ratio was also investigated in this study, where we showed increasing $\text{\Oh}_p$ 
leads to a reduction of the maximum jet height. Additionally, a thicker jet was observed due to increased extensional viscosity in the EVP fluid.

In summary, this study 
investigates the bubble-bursting dynamics in an elasto-viscoplastic fluid, elucidating the nuanced effects of elastic stress relaxation and plasticity across varying Deborah and plasto-capillary numbers. This investigation is particularly pertinent to industrial scenarios involving bubble interactions with complex fluids' free surfaces. Our findings, bridging numerical simulations with potential experimental validations~\citep{rodriguez2023,ji2023}, could further enhance the understanding of initial shape variations and bubble buoyancy in different fluids~\citep{deoclecio2023drop}. While this study primarily focuses on elastic stress relaxation and yield stress effects, future work could incorporate additional complexities like the shear-dependent behaviour of EVP fluids for a more exhaustive understanding~\citep{Brown2014, RostiEVP}. Overall, this research sheds light on the critical interplay of surface tension, elasticity, and plasticity in shaping the bubble bursting phenomena in complex fluidic environments.

\appendix
\section{Non-dimensionalization of governing equations}\label{appendix:non_dimensional_ge}

The governing equations for an incompressible fluid is

\begin{align} \label{eqn:continuity_D}
    \nabla\cdot\boldsymbol{u}&=0\,,\\
    \label{eqn:momentum_D} \rho_l \left(\frac{\partial \boldsymbol{u}}{\partial t} + \boldsymbol{\nabla\cdot}\left(\boldsymbol{uu}\right)\right) &= -\boldsymbol{\nabla} p + \boldsymbol{\nabla\cdot}\left(\boldsymbol{\tau_s} + \boldsymbol{\tau_p} \right) + \rho_l \boldsymbol{g}\,.
\end{align}

\noindent where $\rho_l$ is the density of the fluid, $\boldsymbol{u}$ is the velocity vector and $t$, $p$ corresponds to the time and pressure, respectively. The deviatoric viscous stress tensor and the polymeric stress tensor are represented by~$\boldsymbol{\tau_s}$ and~$\boldsymbol{\tau_p}$, respectively and $g$ corresponds to the acceleration due to gravity. The deviatoric viscous stress tensor is expressed as

\begin{align} \label{eqn:viscous_stress_tensor}
    \boldsymbol{\tau_s} = 2 \mu_s \boldsymbol{\mathcal{D}}\,,
\end{align}

\noindent where $\mu_s$ is the solvent viscosity and $\boldsymbol{\mathcal{D}}$ is the deformation-rate tensor.

According to the thermodynamically consistent constitutive relationship proposed by~\cite{saramito2007} for elasto-viscoplastic fluid, the polymeric stress tensor evolves following the partial differential equation given by:

\begin{align} \label{eqn:polymer_stress_evolution_D}
    \lambda\, \boldsymbol{\overset{\nabla}{\tau}_p} + \mathrm{max}\left(\frac{\|\boldsymbol{\tau_d}\|-\tau_y}{\|\boldsymbol{\tau_d}\|},0 \right )\boldsymbol{\tau_p} = 2\mu_p\, \boldsymbol{\mathcal{D}}\,.
\end{align}

\noindent where, $\lambda$ indicates the relaxation time of the polymers and $\mu_p$, $\tau_y$ corresponds to the polymer viscosity and yield stress of the fluid, respectively. Here, $\boldsymbol{\overset{\nabla}{\tau}_p}$~is the upper-convected time derivative and $\|\boldsymbol{\tau_d}\|$~is the second invariant of the deviatoric part of the polymeric stress tensor $\boldsymbol{\tau_p}$.

For our simulations, we consider the inertia-capillary velocity $(V_\gamma)$, initial bubble radius $(R_0)$, inertial-capillary time $(T_\gamma)$ and the capillary stress $(p_\gamma)$ to non-dimensionalize the equations~(\ref{eqn:continuity_D}--\ref{eqn:polymer_stress_evolution_D}) resulting in non-dimensional equations~(\ref{eqn:continuity_nonD},\ref{eqn:momentum_nonD},\ref{eqn:polymer_stress_evolution}). The above-described quantities are defined as:
\begin{align}
    V_\gamma &= \sqrt{\frac{\gamma}{\rho_l R_0}}\,,\\
    T_\gamma &= \frac{R_0}{V_\gamma} = \sqrt{\frac{\rho_l R_0^3}{\gamma}}\,,\\
    p_\gamma &= \frac{\gamma}{R_0}\,.
\end{align}

The set of equations~(\ref{eqn:continuity_D}-- \ref{eqn:viscous_stress_tensor}) are also solved for the gas phase with~density $\rho_g$, viscosity $\mu_g$ and \hlrev{no extra-stress tensor ($\boldsymbol{\tau_p} = \bold{0}$)}.

\section{Log-conformation approach}\label{appendix:log_conformation}

In the present study, the effect of elasticity and plasticity embedded in an elasto-viscoplastic fluid is modelled with the constitutive equation proposed by~\cite{saramito2007} as

\begin{align} \label{eqn:polymer_evolution_saramito}
    \text{\De}\, \left[ \frac{\partial \boldsymbol{\tau_p}}{\partial t} + \left(\boldsymbol{u\cdot\nabla}\right)\boldsymbol{\tau_p} - \boldsymbol{\tau_p\cdot\nabla u} - \left(\boldsymbol{\nabla u}\right)^T\boldsymbol{\cdot}\boldsymbol{\tau_p} \right] + \mathrm{max}\left(\frac{\|\boldsymbol{\tau_d}\|-\mathcal{J}}{\|\boldsymbol{\tau_d}\|},0 \right )\boldsymbol{\tau_p} = 2\text{\Oh}_p\, \boldsymbol{\mathcal{D}}\,.
\end{align}

In the above model, the unyielded fluid behaves as a Kelvin-Voigt solid whereas the yielded fluid flows as an Oldroyd-B viscoelastic fluid. The polymeric stress can be written in conformation tensor form as

\begin{equation}
	\label{eqn:stressStrain}
	\boldsymbol{\tau_p} = \frac{\mu_p}{\lambda} \left(\boldsymbol{A} - \boldsymbol{I}\right)
\end{equation}

\noindent where, $\boldsymbol{A}$ corresponds to the conformation tensor, an order parameter that keeps track of the stretching of the polymers \citep{snoeijer2020, Stone2023Note}. Filling in equation~\eqref{eqn:stressStrain}, \eqref{eqn:polymer_evolution_saramito} can be re-written as

\begin{align} \label{eqn:polymer_evolution_conformation_tensor}
    \frac{\partial \boldsymbol{A}}{\partial t} + \left(\boldsymbol{u\cdot\nabla}\right)\boldsymbol{A} - \boldsymbol{A\cdot}\left(\boldsymbol{\nabla u}\right) - \left(\boldsymbol{\nabla u}\right)^T\boldsymbol{\cdot}\boldsymbol{A} = -\frac{K}{\text{\De}}\left(\boldsymbol{A} - \boldsymbol{I}\right)\,,
\end{align}

\noindent with the switch-term $K = \mathrm{max}\left((\|\boldsymbol{\tau_d}\|-\mathcal{J})/\|\boldsymbol{\tau_d}\|,0 \right )$.

In order to simulate high Deborah numbers, we use the log conformation formulation~\citep{fattal2004} to preserve positive-definiteness of $\boldsymbol{A}$ and alleviate the high-Weissenberg-number problem (HWNP).

In log conformation formulation, the components of $\boldsymbol{A}$ in equation~\eqref{eqn:polymer_evolution_conformation_tensor} is solved using the split scheme approach similar to that of ~\cite{hao2007} in the basis of a logarithm of the conformation tensor as $\boldsymbol{\psi} = \mathrm{log}\,\boldsymbol{A}$, since $\boldsymbol{A}$ is positive definite. The velocity gradient tensor $\nabla \boldsymbol{u}$ is decomposed into two anti-symmetric tensors $\boldsymbol{\Omega}$ and $\boldsymbol{N}$, and a symmetric tensor $\boldsymbol{B}$ such that $\nabla \boldsymbol{u} = \boldsymbol{\Omega} + \boldsymbol{N}\boldsymbol{A}^{-1} + \boldsymbol{B}$. Hence, equation~\eqref{eqn:polymer_evolution_conformation_tensor} in the log conformation tensor formulation reads

\begin{align}
    \label{eqn:log_conform}
    \frac{\partial \boldsymbol{\psi}}{\partial t} + \left(\boldsymbol{u\cdot\nabla}\right)\boldsymbol{\psi} - \left(\boldsymbol{\Omega} \boldsymbol{\psi} - \boldsymbol{\psi} \boldsymbol{\Omega}\right) - 2\boldsymbol{B} = \frac{K}{\text{\De}}\left(\mathrm{exp}\left(-\boldsymbol{\psi}\right) - \boldsymbol{I}\right)\,.
\end{align}

\noindent \hlrev{The extra stress tensor~$\boldsymbol{\tau_p}$ evolution is solved explicitly in time, with its divergence incorporated as a pressure-gradient-like source term in the momentum equation~\eqref{eqn:momentum_D}. The explicit time-stepping for both~$\boldsymbol{\tau_p}$ evolution and surface tension necessitates a sufficiently small time step to maintain numerical stability. For the cases we have simulated, the capillary time step restriction (discussed in~\S~\ref{subsec: sim_setup} and \citet{popinet2009}) proves limiting.} The readers are referred to~\cite{lopez2019} for the corresponding implementation details in Basilisk C. \hlrev{Our implementation of the EVP model in Basilisk C can be found at~\citet{ariGithub}.}

\section{Grid-dependence study}\label{appendix:grid_independence}
\begin{figure}
\centering
\resizebox*{0.45\linewidth}{!}{\includegraphics{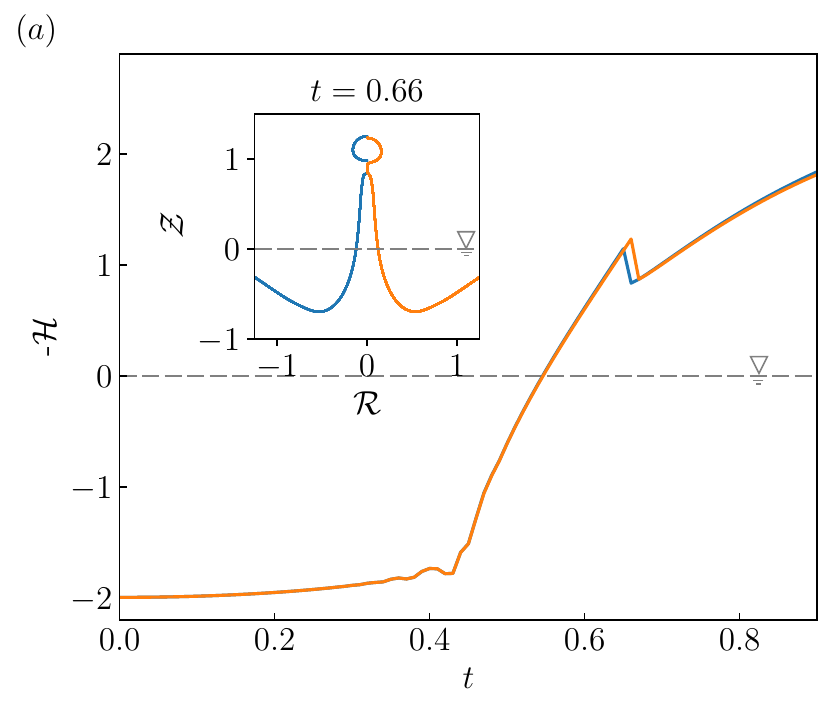}}
\resizebox*{0.46\linewidth}{!}{\includegraphics{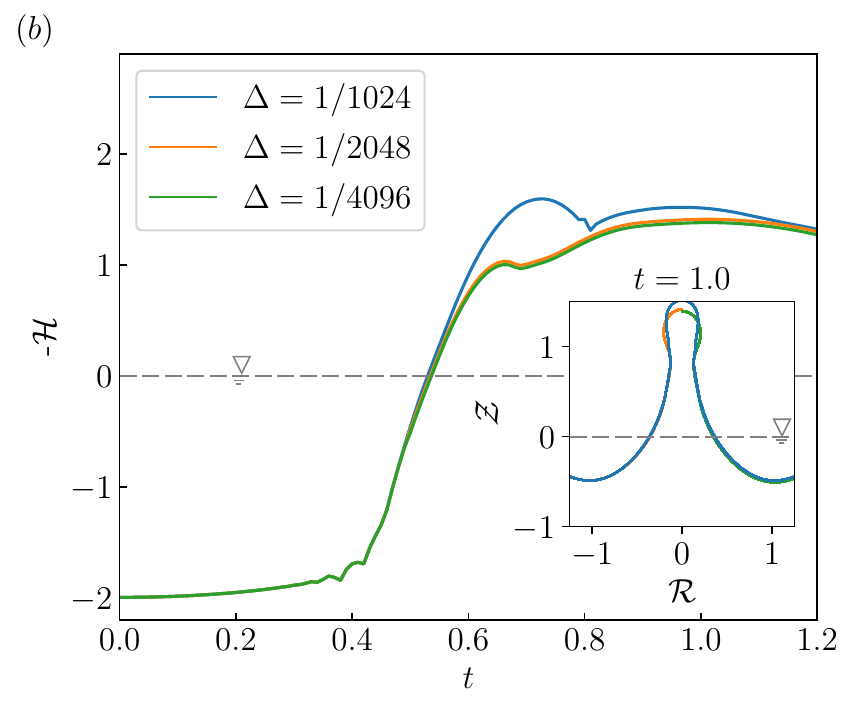}}
\caption[]{Variation of the interface height at the axis for~$(a)$~$\mathcal{J}=0.01,\,\text{\De}=0.01$, $(b)$~$\mathcal{J}=1.0,\,\text{\De}=16$ at different grid resolutions. Inset shows the liquid interface at the $(a)$~drop formation time and $(b)$~at around the maximum jet height.
}
\label{app_fig:grid_dependence}
\end{figure}

Basilisk employs a quadtree grid system that enables resolutions in powers of 2, denoted as $2^n$. \hlrev{Given our chosen square computational domain of size $8R_0$, the number of uniform grid points across the initial bubble radius $R_0$ is $2^{n-3}$ and thereby the minimum cell size is $\Delta = 1/2^{n-3}$.} In our study, we implement three different resolution levels: $\Delta = 1/2^{n-3}, 1/2^{n-4},$ and $1/2^{n-5}$, depending on the proximity to the bubble. Specifically, we use a grid resolution of $\Delta = 1/2^{n-3}$ very close to the bubble $(r<1.28 R_0)$ and $1/2^{n-5}$ near the boundary. Figure~\ref{app_fig:grid_dependence} demonstrates the grid size independence of the bubble bursting results in terms of jet development.

From figure~\ref{app_fig:grid_dependence}, we see that the adaptive scheme yields similar results for $n=13,14$ at low $\text{\De}$. For high values of elasticity, similar jet development is observed for $n=14$ and $15$. \hlrev{It should be noted that the droplet break-up is mesh-size dependent~(\textit{i.e.} a  droplet is formed when the thickness of the jet is smaller than the grid size)} and hence, we have focused on a qualitative description of droplet break-up time rather than a quantitative study. Furthermore, we have observed that the difference in droplet break-up time does not significantly impact later jet development process, leading to grid-converged results. For higher $\text{\De}$, given that the Oldroyd-B model allows for the buildup of infinite stress, we find that a sufficiently fine grid is necessary to accurately resolve the elastic stresses, which are predominantly concentrated in the axial region of the jet.

\section{Comparison of Bingham model and EVP model at viscoplastic limit}\label{appendix:Bingham_EVP}

The constitutive equation for a Bingham viscoplastic fluid (in non-dimensional form) is given by
\begin{equation} \label{eqn:Bingham_model}
    \begin{aligned}
    \boldsymbol{\mathcal{D}} &= \bold{0} \hspace{3.2cm} \|\boldsymbol{\tau}\| < \mathcal{J}\,,\\
    \boldsymbol{\tau} &= 2\left(\frac{\mathcal{J}}{2\|\boldsymbol{\mathcal{D}}\|} + \text{\Oh} \right)\boldsymbol{\mathcal{D}} \hspace{0.5cm} \|\boldsymbol{\tau}\| \ge  \mathcal{J}\,.
\end{aligned}
\end{equation}

\hlrev{With the introduction of a `large' viscous regularisation parameter $\text{\Oh}_{max}$ in equation~\eqref{eqn:Bingham_model} for the unyielded state}, the regularised Bingham model reads

\begin{equation}
    \label{eqn:reg_Bingham_model}
    \boldsymbol{\tau} = 2\, \mathrm{min} \left(\text{\Oh}_{max}\,, \frac{\mathcal{J}}{2\|\boldsymbol{\mathcal{D}}\|} + \text{\Oh}\right) \boldsymbol{\mathcal{D}}\,.
\end{equation}

Nevertheless, the stress in the yielded region of viscoplastic liquid is

\begin{equation} \label{eqn:Bingham_yield}
	 \boldsymbol{\tau} = 2\left(\frac{\mathcal{J}}{2\|\boldsymbol{\mathcal{D}}\|} + \text{\Oh}\right)\boldsymbol{\mathcal{D}} = \frac{\mathcal{J}}{\|\boldsymbol{\mathcal{D}}\|}\boldsymbol{\mathcal{D}} + 2\text{\Oh} \boldsymbol{\mathcal{D}}\,.
\end{equation}

However for the EVP model in this study, towards viscoplastic limit ($\text{\De} \rightarrow 0$) and specifically~$\text{\De}=0$ from equations~\eqref{eqn:SolventRheo} and~\eqref{eqn:polymer_stress_evolution}, we obtain

\begin{align}
    \boldsymbol{\tau} = \boldsymbol{\tau_p} + \boldsymbol{\tau_s} &= 2\,\left(\frac{\text{\Oh}_p}{K} + \text{\Oh}_s\right)\boldsymbol{\mathcal{D}}\,,\\
    &=   2(1-\beta)\text{\Oh}\boldsymbol{\mathcal{D}}\left(\frac{1-K}{K}\right) + 2\text{\Oh}\boldsymbol{\mathcal{D}}\,,\\
    \label{eqn:EVP_Bingham}&= \frac{2(1-\beta)\text{\Oh}\mathcal{J}}{\|\boldsymbol{\tau_d}\|-\mathcal{J}}\boldsymbol{\mathcal{D}}+2\text{\Oh}\boldsymbol{\mathcal{D}}\,,
\end{align}
where definition of $K$ from equation~\eqref{eqn:Aconform} for the yielded state of EVP fluid is used in equation~\eqref{eqn:EVP_Bingham}.

From equations~\eqref{eqn:Bingham_yield} and~\eqref{eqn:EVP_Bingham}, we observe that the regularised Bingham model and EVP model can be equivalent if
\begin{equation}
    \frac{1}{\|\boldsymbol{\mathcal{D}}\|} = \frac{2(1-\beta)\text{\Oh}}{\|\boldsymbol{\tau_d}\|-\mathcal{J}}\,.
\end{equation}
The above equation holds for~$\beta = 0$ and if the value of $\beta = \text{\Oh}_s/\text{\Oh} > 0$, then the effective yielded viscosity in the EVP fluid is lower than that in the Bingham viscoplastic model.

\section{Energy-budget calculation}
\label{appendix:energy_budget}

The formulation for different modes of energy transfer is presented here. Similar energy-transfer mode calculations have been employed in prior works, such as~\cite{ramirez2020} and~\cite{vatsal2021}, for evaluating energy budgets in scenarios involving colliding droplets and bubble bursting, respectively. In our study, we extend this methodology to elasto-viscoplastic liquids. Building upon the work by~\cite{snoeijer2020} in energy analysis for Oldroyd-B viscoelastic fluids, we adapt the formulation to elasto-viscoplastic fluids.

From equation~\eqref{eqn:momentum_nonD}, the rate of change of kinetic energy can be written as:
\begin{align}
    \frac{1}{2}\frac{\partial \rho u^2}{\partial t} + \nabla\cdot\left[ \left( \frac{\rho u^2}{2} + p \right)\boldsymbol{u} - 2\mu\boldsymbol{\mathcal{D}}\cdot\boldsymbol{u} - \boldsymbol{\tau_p}\cdot\boldsymbol{u} \right] = -\epsilon_s -\xi_p\,.
\end{align}

The terms in the square braces correspond to the energy flux terms and the last two terms in the above equation corresponds to work done due to solvent viscosity and polymers, respectively and is written as,
\begin{align}
    \epsilon_s &= 2\text{\Oh}_s \boldsymbol{\mathcal{D}}\colon\boldsymbol{\mathcal{D}}\,, \\
    \xi_p &= \boldsymbol{\tau_p} \colon \boldsymbol{\mathcal{D}}\,.
\end{align}

The work done due to polymers can be further split into
\begin{align}
    \boldsymbol{\tau_p} \colon \boldsymbol{\mathcal{D}} = \frac{dW}{dt} + \epsilon_p\,,
\end{align}
where $W$ is the elastic energy density and $\epsilon_p$ is the dissipation of energy due to polymers.
It should be noted that the elastic energy is associated to the polymer elongation and thereby is a function of~$\mathrm{tr}(\boldsymbol{A})$. Similarly, the relaxation of elastic stresses (thereby relaxation of conformation tensor~$\boldsymbol{\overset{\nabla}A}$) gives rise to polymer dissipation~$\epsilon_p$. Here, $\boldsymbol{\overset{\nabla}A}$~corresponds to the terms in the square braces of equation~\eqref{eqn:polymer_evolution_conformation_tensor}.

For an Oldroyd-B model, starting from the elastic stress evolution equation~\eqref{eqn:polymer_evolution_conformation_tensor} we have
\begin{align}
    \overset{\nabla}{\boldsymbol{A}} = -\frac{K}{\text{\De}}\left(\boldsymbol{A} - \boldsymbol{I}\right)\,,
\end{align}
with $K = 1$. Taking the trace of the constitutive equation, we arrive at the following expressions
\begin{align}
    W &= \frac{\text{\Oh}_p}{2\text{\De}}\left(\mathrm{tr}\left(\boldsymbol{A} - \boldsymbol{I}\right)\right) = \frac{Ec}{2}\left(\mathrm{tr}\left(\boldsymbol{A} - \boldsymbol{I}\right)\right) \,,\\
    \epsilon_p &= \frac{W}{\text{\De}} = \frac{Oh_p}{De^2}\left(\mathrm{tr}\left(\boldsymbol{A} - \boldsymbol{I}\right)\right)\,,
\end{align}
for elastic energy and polymer dissipation, respectively.

The key distinction between the viscoelastic Oldroyd-B model and the~\cite{saramito2007} model lies in the presence of the switch term~$K$ (which mimicks the effect of plasticity). It is straightforward to demonstrate that this switch term does not affect the elastic energy of the EVP fluid. The above conclusion is based on the definition that, before yielding of the elastoviscoplastic (EVP) fluid, it behaves as a Kelvin-Voigt viscoelastic solid, which inherently includes the neo-Hookean spring. However, once the fluid has yielded, the polymer dissipation $\epsilon_p = f(\boldsymbol{\overset{\nabla}{A}})$ is identical to that in the Oldroyd-B model. The expressions for the elastic energy and polymer dissipation in an EVP are respectively given by,
\begin{align}
    W &= \frac{\text{\Oh}_p}{2\text{\De}}\left(\mathrm{tr}\left(\boldsymbol{A} - \boldsymbol{I}\right)\right) = \frac{Ec}{2}\left(\mathrm{tr}\left(\boldsymbol{A} - \boldsymbol{I}\right)\right)\,,\\
    \label{eq:De2}
    \epsilon_p &= \frac{K W}{\text{\De}} = \frac{K Oh_p}{De^2}\left(\mathrm{tr}\left(\boldsymbol{A} - \boldsymbol{I}\right)\right)\,.
\end{align}

In~\S\ref{subsec:energy_budget} we analyze the temporal evolution of energy within the computational domain, following a similar approach as used in~\cite{vatsal2021}. The kinetic and the surface energy of the liquid are respectively given by

\begin{align}
\label{eqn:kinetic_energy}
    E_k &= \frac{1}{2} \int_{\Omega} \|\boldsymbol{u}\|^2\, {\rm d}\Omega\,,\\
    E_s &= \int_{\Gamma} {\rm d}\Gamma - \int_{\zeta} {\rm d}\zeta\,.
\end{align}
It should be noted that the energies are normalised by the surface energy~$\gamma R_0^2$. The integrals for the energy in the fluid are evaluated over the volume~($\Omega$) and the surface~($\Gamma$) of the largest liquid continuum in the computational domain and does not encompass the drops (which are constituted under a separate term~$E_m$). Furthermore, the state of the liquid pool, characterized by a flat free surface $\zeta$ ($\mathcal{Z}=0$), is considered as the reference condition.

The dissipation due to the solvent part of the EVP fluid is given by
\begin{align}
    E_d^s = 2 \text{\Oh}_s \int_{t} \left( \int_{\Omega} \| \boldsymbol{\mathcal{D}}\|^2\, {\rm d}\Omega \right) {\rm d}t\,.
\end{align}

The elastic energy and the polymer dissipation are quantified as
\begin{align}
    E_e &= \int_{\Omega} W {\rm d}\Omega\,,\\
    E_d^{p} &= \int_{t} \left(\int_{\Omega} \epsilon_p\, {\rm d}\Omega \right) {\rm d}t\,. \label{eqn:polymer_dissipation}
\end{align}

All other forms of energy is constituted as
\begin{align}
\label{eqn:Em}
    E_m = E_k^{\rm Drops} + E_s^{\rm Drops} + E_d^{\text{\Oh},\, {\rm Drops}} + E_p^{\rm Drops} + \int_{\Omega + \Omega_d} \mathcal{B}o \mathcal{Z}\, d(\Omega + \Omega_d) + E_g\,.
\end{align}

In equation~\eqref{eqn:Em} the superscript `\rm{Drops}' corresponds to the ejected drops with the corresponding volume denoted by~$\Omega_d$. Further, the terms respectively correspond to the kinetic energy, surface energy, viscous dissipation and polymer work in the drops. The volume integrals are performed in the same way as in equations~(\ref{eqn:kinetic_energy}--\ref{eqn:polymer_dissipation}) except that the integrals are performed over~$\Omega_d$. The last two terms correspond to the gravitational potential for the fluid and the total energy stored in the gas phase, respectively.

The total energy in the gas phase occupying a volume~$\Omega_g$, as outlined in~\cite{vatsal2021}, is written as
\begin{align}
    E_g = \rho_r \int_{\Omega_g} \left( \frac{\|\boldsymbol{u}\|^2}{2} + \mathcal{B}o \mathcal{Z} \right) {\rm d}\Omega_g + 2\mu_r \text{\Oh} \int_{t} \left( \int_{\Omega_g} \| \boldsymbol{\mathcal{D}} \|^2\, {\rm d}\Omega_g \right) {\rm d}t\,.
\end{align}

\begin{figure}
\centering
\resizebox*{0.45\linewidth}{!}{\includegraphics{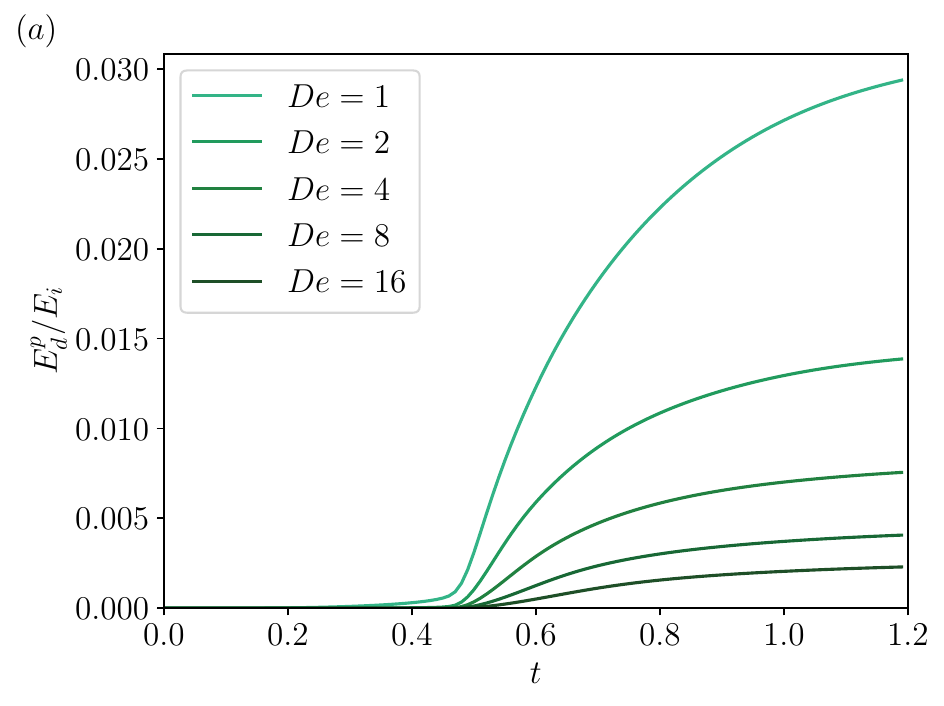}}
\resizebox*{0.45\linewidth}{!}{\includegraphics{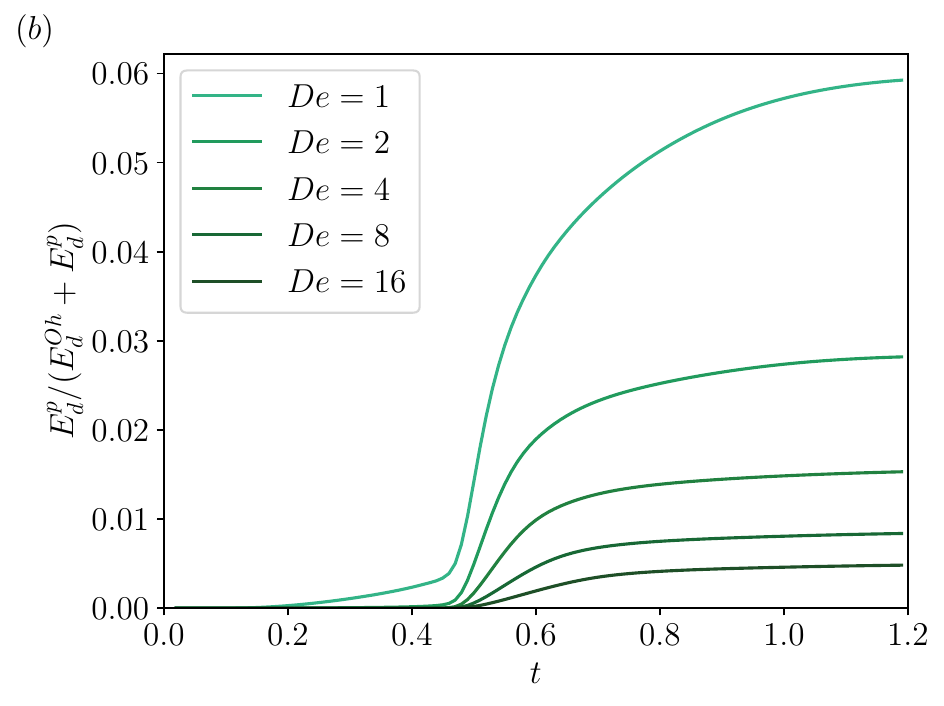}}
\resizebox*{0.45\linewidth}{!}{\includegraphics{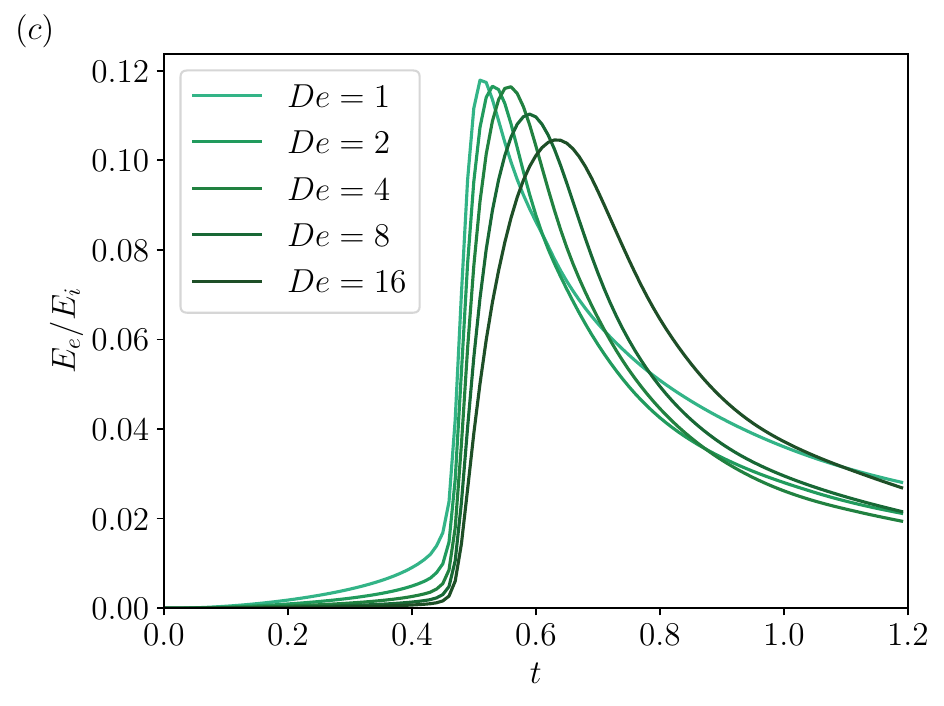}}
\caption[]{Temporal variation of $(a)$~polymeric dissipation, $(b)$~ratio of polymeric dissipation to viscous and polymeric dissipation and $(c)$~elastic energy, normalised with the initial energy for different~$\text{\De}$ in the visco-elasto-capillary regime at $\mathcal{J}=0.1$.
}
\label{app_fig:visco_elasto_capillary_1}
\end{figure}

\begin{figure}
\centering
\resizebox*{0.45\linewidth}{!}{\includegraphics{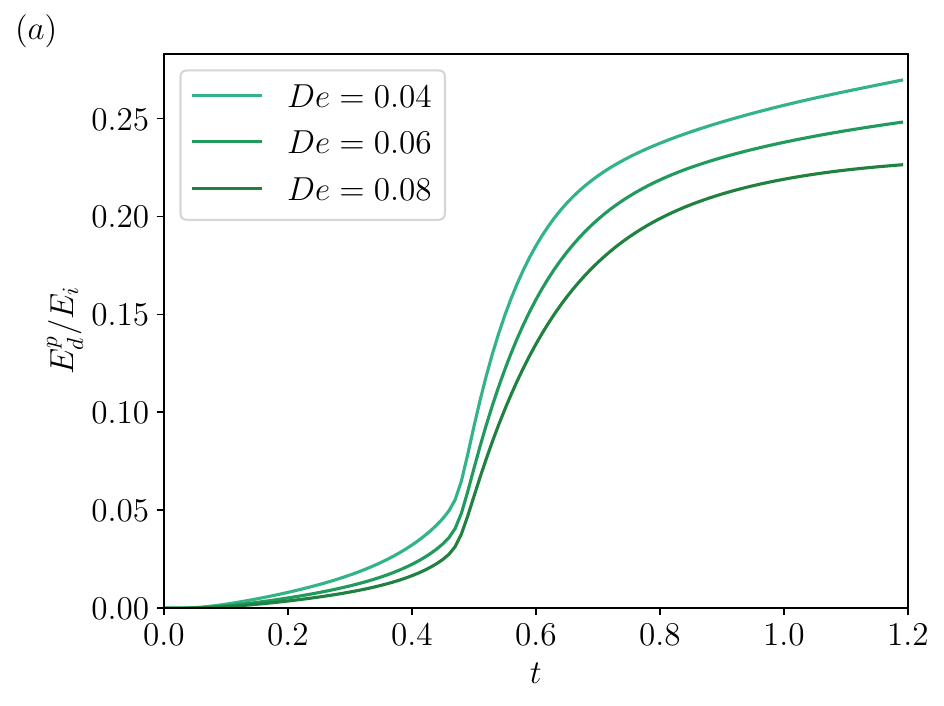}}
\resizebox*{0.45\linewidth}{!}{\includegraphics{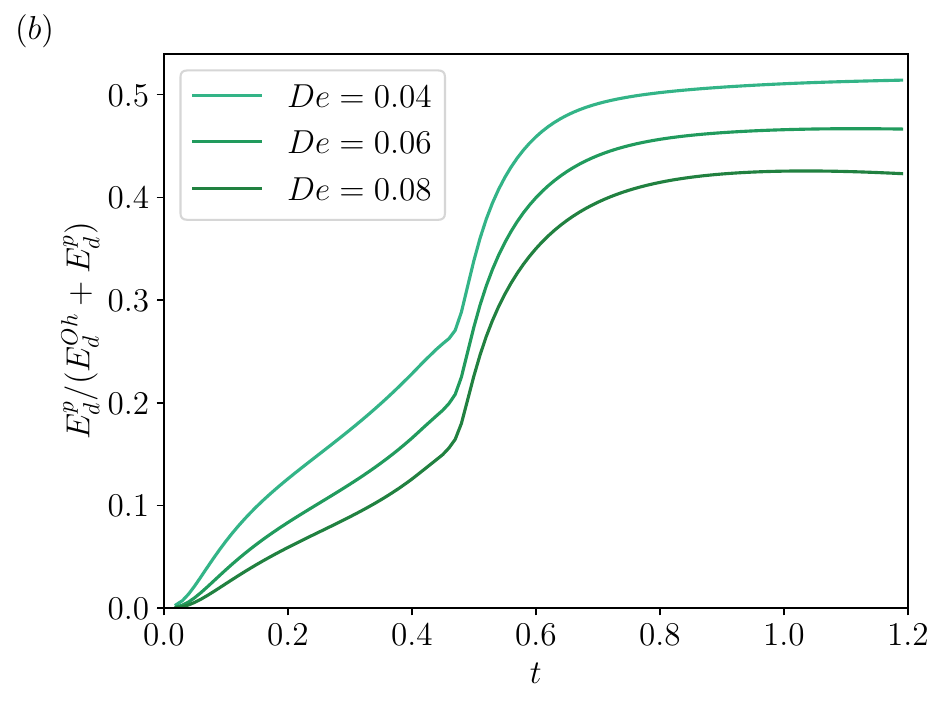}}
\resizebox*{0.45\linewidth}{!}{\includegraphics{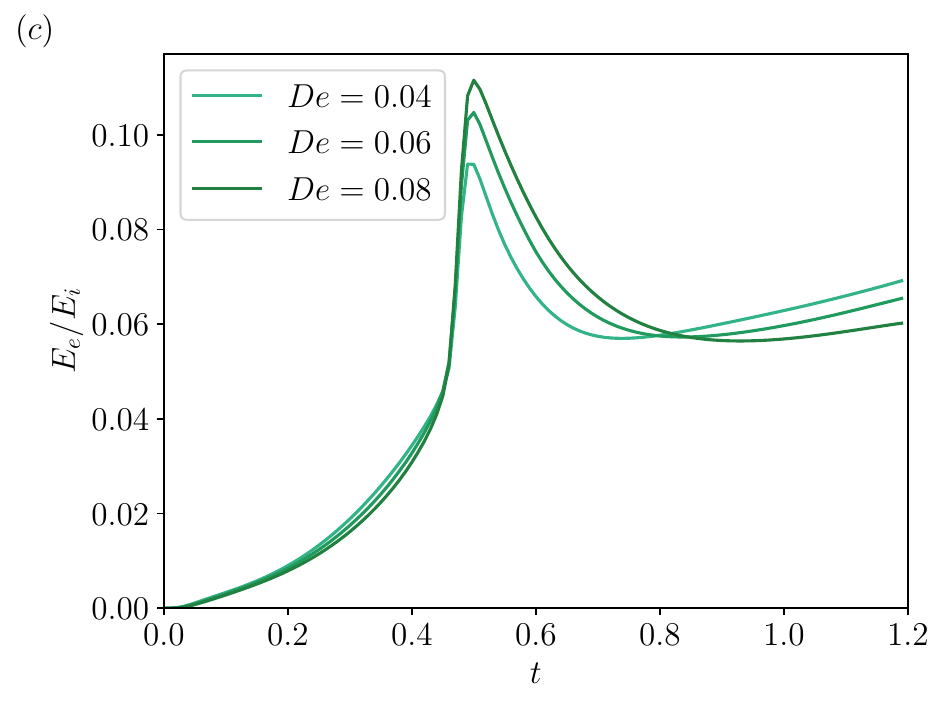}}
\caption[]{Temporal variation of $(a)$~polymeric dissipation, $(b)$~ratio of polymeric dissipation to viscous and polymeric dissipation and $(c)$~elastic energy, normalised with the initial energy for low~$\text{\De}$ and~$\mathcal{J}=0.1$.
}
\label{app_fig:visco_elasto_capillary_2}
\end{figure}

\section{Different jet growth at low~$\text{\De}$ and high~$\text{\De}$} \label{appendix:jet_growth}

From figure~\ref{fig_height_De}$a$, we observed that the jet height increases with~$\text{\De}$ for~$\text{\De}\sim \mathcal{O}(1)$ in the visco-elasto-capillary regime. In this section, we plot the temporal evolution of corresponding polymeric dissipation, ratio of polymeric dissipation to the total dissipation contribution from viscosity and polymers and elastic energy in figure~\ref{app_fig:visco_elasto_capillary_1}$a$--$c$, respectively. From figure~\ref{app_fig:visco_elasto_capillary_1}$a$, we observe that the polymeric dissipation during the jet development phase constitutes to only around 3\% of the total initial surface energy for~$\text{\De}=1$ and decreases with increase in~$\text{\De}$. Viscous dissipation increases with~$\text{\De}$ such that, the ratio of the polymeric dissipation to the total viscous and polymeric dissipation decreases with~$\text{\De}$, as observed in figure~\ref{app_fig:visco_elasto_capillary_1}$b$. In the considered range of~\text{\De}, the polymeric dissipation is less than 6\% in comparison to the total dissipation of energy in the system. Further, the elastic energy in the EVP fluid is more pronounced closer to $t\approx 0.5-0.6$ as shown in figure~\ref{app_fig:visco_elasto_capillary_1}$c$, where the merged capillary waves result in jet formation. The corresponding decrease of elastic energy with increase in~$\text{\De}$ is due to lower elasto-capillary number. The elastic stresses are concentrated at a smaller region in the domain and hence, overall the elastic energy in the EVP fluid (because of unyielded region also) is decreasing with~$\text{\De}$ at jet initiation. (Note that the unyielded region behaves as an elastic solid with low elastic modulus.) The combined effects of increasing viscous dissipation and decreasing polymer dissipation and elastic energy at jet initiation contributes to an increase in kinetic energy of the jet and thereby results in higher jet growth with respect to~$\text{\De}$ in the visco-elasto-capillary regime.

Whereas in figure~\ref{fig_height_De}$b$, the jet height is shown to decrease with~$\text{\De}$ of the fluid for~$\text{\De}\sim \mathcal{O}(0.01)$. In order to elucidate the different behaviour of jet growth, the corresponding temporal evolution of polymeric dissipation is plotted in figure~\ref{app_fig:visco_elasto_capillary_2}$a$. Here, $\text{\De}=0.01,0.02$ is not considered as they result in drop formation. The polymeric dissipation shows a decrease with increase in~$\text{\De}$, similar to the observation in figure~\ref{app_fig:visco_elasto_capillary_1}$a$. However, the magnitude of polymeric dissipation is considerably higher in comparison to that observed for~$\text{\De}\sim \mathcal{O}(1)$. This is due to higher elastocapillary number and thereby the elastic energy is larger with significant polymer dissipation~(due to lower~$\text{\De}$). Further, the ratio of polymeric dissipation to total dissipation is plotted in figure~\ref{app_fig:visco_elasto_capillary_2}$b$, which indicates that the polymeric dissipation contributes to around 40\% of the total dissipation of energy. This is because of the larger yielded region at~$\text{\De}\sim\mathcal{O}(0.01)$, where polymeric activity is pronounced. The variation of elastic energy with respect to~$\text{\De}$ is shown in figure~\ref{app_fig:visco_elasto_capillary_2}$c$. The increasing elastic energy at the time of jet initiation~($t\approx 0.5$) is due to the increased storage of elastic energy both in the yielded region (where elastic stress is relaxing slower) and relatively larger unyielded region that exhibits solid behaviour~(refer figures~\ref{fig_index_plot_De_1}$b$, \ref{fig_index_plot_J}$b$). The increase in elastic energy results in strain-rate hardening and thereby reduces the jet growth. Note that this variation in the elastic energy also affects the initiation kinetic energy of the jet and thereby the maximum initiation velocity of the jet. Further, the jet development is dictated by the capillary stresses and storage and relaxation of elastic stresses in the jet. Overall, the observation of lower maximum jet height with respect to~$\text{\De}$ for $\text{\De}\sim\mathcal{O}(0.01)$ is related to the elastic energy component of the EVP fluid. 

\backsection[Acknowledgements]{The simulations were run with the computational resources provided by the National Academic Infrastructure for Supercomputing in Sweden (NAISS).}

\backsection[Funding]{This work is supported by the funding provided by the European Research Council grant no.~"2021-StG-852529, MUCUS" and the Swedish Research Council through grant No 2021-04820. RV acknowledges the ERC grant no.~"2021-CoG-101043998, DEEPCONTROL". VS acknowledges the financial support from the Physics of Fluids at the University of Twente under the FIP-II project, sponsored by NWO and Canon. MJ acknowledges the ERC grant no.~"2023-StG-101117025, FluMAB" and funding by the
NWO (Dutch Research Council) under the MIST (MItigation STrategies for Airborne Infection Control) program. Views and opinions expressed are however those of the author(s) only and do not necessarily reflect those of the European Union or the European Research Council. Neither the European Union nor the granting authority can be held responsible for them.}

\backsection[Declaration of interests]{The authors report no conflict of interest.}

\backsection[Data availability statement]{The data that support the findings of this study will be openly available on GitHub--\cite{ariGithub} upon publication.}

\backsection[Author ORCIDs]
{\\ A. G. Balasubramanian \href{https://orcid.org/0000-0002-0906-3687}{https://orcid.org/0000-0002-0906-3687}\\ V. Sanjay \href{https://orcid.org/0000-0002-4293-6099}{https://orcid.org/0000-0002-4293-6099} \\ M. Jalaal \href{https://orcid.org/0000-0002-5654-8505}{https://orcid.org/0000-0002-5654-8505} \\ R. Vinuesa \href{https://orcid.org/0000-0001-6570-5499}{https://orcid.org/0000-0001-6570-5499} \\ O. Tammisola \href{https://orcid.org/0000-0003-4317-1726}{https://orcid.org/0000-0003-4317-1726}}

\bibliographystyle{jfm}
\bibliography{literature}

\end{document}